\documentclass[a4paper,11pt]{article}
\pdfoutput=1 

\usepackage{jcappub} 

\usepackage[T1]{fontenc} 

\usepackage{amsmath, amsfonts, mathtools, amsthm, amssymb, mathtools}
\usepackage{subfig}
\usepackage{tikz-cd, tikz}
\usepackage{graphicx}
\usepackage{float}
\usepackage{wrapfig}
\usepackage{algorithm}
\usepackage{listings}
\usepackage[noend]{algpseudocode}
\usepackage{appendix}
\usepackage{listings}

\title{\boldmath Fluctuating Dark Energy and the Luminosity Distance}

\author[]{Casper J.G. Vedder,}
\author[]{Enis Belgacem,}
\author[]{Nora Elisa Chisari,}
\author[]{Tomislav Prokopec}

\affiliation[]{Institute for Theoretical Physics, Utrecht University, Princetonplein 5, 3584 CC Utrecht, The Netherlands.}

\emailAdd{c.j.g.vedder@students.uu.nl}
\emailAdd{e.belgacem@uu.nl}
\emailAdd{n.e.chisari@uu.nl}
\emailAdd{t.prokopec@uu.nl}

\abstract{The origin of dark energy driving the accelerated expansion of the universe is still mysterious. We explore the possibility that dark energy fluctuates, resulting in spatial correlations. Due to these fluctuations, the Hubble rate itself becomes a fluctuating quantity. We discuss the effect this has on measurements of type Ia supernovae, which are used to constrain the luminosity distance. We show that the luminosity distance is affected by spatial correlations in several ways. First, the luminosity distance becomes dressed by the fluctuations, thereby differing from standard $\Lambda$CDM. Second, angular correlations become visible in the two-point correlation function of the luminosity distance. To investigate the latter we construct the angular power spectrum of luminosity distance fluctuations. We then perform a forecast for two supernova surveys, the ongoing Dark Energy Survey (DES) and the upcoming Legacy Survey of Space and Time (LSST), and compare this effect with relativistic lensing effects from perturbed $\Lambda$CDM. We find that the signal can rise above the lensing effects and that LSST could test this effect for a large part of the parameter space. As an example, a specific realisation of such a scenario is that quantum fluctuations of some field in the early universe imprint spatial correlations with a predictable form in the dark energy density today. In this case, the Hubble rate fluctuates due to the intrinsic quantum nature of the dark energy density field. We study whether the signal of this specific model would be measurable, and conclude that testing this model with LSST would be challenging. However, taking into account a speed of sound $c_s<1$ of the dark energy fluid can make this model observable.}

\begin{document}
\maketitle
\flushbottom

\section{\label{sec:introduction}Introduction}

Since cosmic acceleration was discovered in 1998 \cite{Riess1998,perlmutter1999}, its physical origin has provided a great challenge to physicists and astronomers alike. The standard cosmological model $\Lambda$CDM, with a cosmological constant $\Lambda$, has so far been the most successful model to explain this acceleration. However, the success of $\Lambda$CDM comes at a cost, as the dark sector (dark energy and dark matter) is physically ill understood. Recent observations have also shown tensions between parameters measured when probed at different scales, most notable are the measurements of the Hubble constant $H_0$, where the tension has now overcome 4$\sigma$ \cite{Verde_2019,Di_Valentino_2021}. Another, somewhat weaker tension has also appeared between the $\sigma_8$ measurements by \textit{Planck} \citep{planck2018} and cosmic shear data from e.g. KiDS \citep{Heymans_2021} or DES \citep{Abbott_2022}, this discrepancy is now about 2.5$\sigma$. 

In light of these problems, considerable work has been done on models of dark energy (DE). Options ranging from modifying gravity to matter condensates 
\cite{austin_2016,tsujikawa2013,baker2013,starobinsky2000}. Predictions usually focus on the dynamic aspect of dark energy. For example, modifying dark energy at late time \citep{Di_Valentino_2016,Di_Valentino_2020,Li_2019,vattis_2019}, or by altering the early universe \citep{Hill_2020,Karwal_2016,Poulin_2019,Jedamzik_2020}. We pursue another route, focusing on the spatial correlations dark energy may exhibit. These correlations naturally arise in some classes of models. An example is the model previously considered by \cite{Glavan2014,Glavan2015,Glavan2016,Glavan2018,Belgacem2021,Belgacem2022}. In this model dark energy arises from quantum fluctuations in the early universe. It has recently been shown that this model, due to the spatial fluctuations in dark energy, can reduce the Hubble tension towards $1 \sigma$ \cite{Belgacem2021}. This is a consequence of the spatial correlations that dark energy exhibits in this model, which lead to a fluctuating Hubble rate.

In the near future telescopes such as the Rubin Observatory, which will deliver the \textit{Legacy Survey of Space and Time}  (LSST) \cite{lsst}, will obtain maps of the sky with unprecedented detail. This will allow us to discriminate in the wide range of models currently available. It is therefore of importance to have predictions of a certain class of models containing these spatial fluctuations.  

To this end we propose a phenomenological model of dark energy with spatial fluctuations, causing the background to fluctuate. This leads to a fluctuating Hubble rate, where at each point the effect of dark energy is such that the Hubble rate obeys the following operator Friedmann equation,
\begin{equation}\label{eq:frdef}
3 M_{P}^{2} \hat{H}^{2}(z, \hat{n})=\rho_{C}(z) \hat{\operatorname{I}}+\hat{\rho}_{Q}(z, \hat{n}).
\end{equation}
Here, ${M}_{P}=2.435\times10^{18} \operatorname{ GeV}/c^2$ is the reduced Planck mass, $\rho_C$ is the energy density of classical matter and $\hat{\rho}_Q$ is the energy density operator containing the dark energy field, $\hat{\operatorname{I}}$ is the identity operator. These operators can be understood as acting on a Hilbert space of quantum states and describe the fluctuating fields. $\hat{H},\rho_c$ and $\hat{\rho}_Q$ are functions of the redshift $z$, the operators also depend on the unit vector on the celestial sphere $\hat{n}$. For the correlations we use a phenomenological power law ansatz, which is described in more detail in section \ref{sec:model}, but in section \ref{sec:quantummodel} we study a simple inflationary model \citep{Glavan2014,Glavan2015,Glavan2016,Glavan2018,Belgacem2021} where a scalar field produces such spatial correlations \citep{Belgacem2021}.

A fluctuating Hubble rate would induce correlations in a wide variety of Large Scale Structure observables. In this paper, we focus on the luminosity distance. Observationally, the luminosity distance is obtained from the `known' luminosity of a standard candle and the observed flux. Theoretically, this distance can be expressed in terms of spacetime geometry. In $\Lambda$CDM this leads to the well known expression in terms of the Hubble rate, ${d}_L(z) = (1+z)\int_0^z dz^\prime{{H}(z^\prime)^{-1}}$ (see e.g. \citep{weinberg2008cosmology}). However, our Hubble rate is a fluctuating quantity determined by (\ref{eq:frdef}), therefore $d_L(z)$ also becomes a fluctuating quantity. Consequently, we turn the luminosity distance into an operator:
\begin{equation}\label{eq:obsdis}
    \hat{d}_L(z,\hat{n}) = (1+z)\int_0^z dz^\prime{\hat{H}^{-1}(z^\prime,\hat{n})}.
\end{equation}
We make predictions for the correlations our model produces and assess its detectability, comparing it with the expected noise and with effects that would be expected to be seen in perturbed $\Lambda$CDM, where these fluctuations come from relativistic effects such as convergence and the Doppler effect.

These relativistic effects on the luminosity distance have been well studied \cite{sasaki1987,Bonvin_2006,Biern_2017}, with some work also having been done on these fluctuations in theories with dynamical dark energy/modified gravity \cite{Garoffolo2021} or in the case of inhomogeneous dark energy \cite{Cooray_2010}. In the latter case the convergence effect of inhomogeneities was studied  for a phenomenological model of dark energy. These papers, however, do not study the effects of a fluctuating Hubble rate but rather the effect of density perturbations between the emitter and the observer. Both effects result in a non zero power spectrum for the luminosity distance fluctuations.

This paper is organised as follows. In section \ref{sec:model} we describe the specifics of our phenomenological model, in section \ref{sec:luminositydistance} we develop the equations needed to make predictions for correlators of the luminosity distance. In section \ref{sec:ps} we then calculate the corresponding angular power spectrum. Section \ref{sec:obs} discusses the forecasting methods, the relativistic effects, and how we obtain our model parameters. Section \ref{sec:results} contains our results and forecasts for this angular power spectrum. Lastly, in section \ref{sec:quantummodel}, we apply our formalism specifically for the previously discussed model where dark energy arises from quantum fluctuations in the early universe.

\section{The model}\label{sec:model}
To develop equations for the luminosity distance, we first need to specify the correlations that we are considering. For our phenomenological model we assume the following equal time ansatz for the correlations:
 \begin{equation}\label{eq: ansatz}
     \langle \hat{\rho}_Q (z,n)\hat{\rho}_Q(z,n^\prime) \rangle =  \langle \hat{\rho}_Q (z) \rangle \langle \hat{\rho}_Q (z) \rangle s(\|\vec{x}-\vec{y}\|).
 \end{equation}
 Here, $\vec{x}$ and $\vec{y}$ are comoving positions associated with the relevant coordinates and $\langle\ldots\rangle$ denotes an ensemble average. As a result of the statistical homogeneity and isotropy, the function $s$ only depends on the relative distance between the coordinates. We consider two forms of the function $s(\|\vec{x}-\vec{y}\|)$, motivated by different hypothesis of how the quantum dark energy density field depends on the fundamental fields that build it. For simplicity, we suppose that the quantum dark energy density field originates from a single fundamental quantum field $\hat{\phi}$.
 
 The first hypothesis we consider can be understood from a theory  where $\hat{\rho}_Q\propto\hat{\phi}^2$. We assume the field $\hat{\phi}$ to be Gaussian, making the density field $\hat{\rho}_Q$ inherently non-Gaussian. Such theories would naturally arise when the corresponding field has a zero vacuum expectation value (vev), which we assume in \textit{Case I}, in which case the energy momentum tensor relates the energy density to $\hat{\phi}^2$. An example of such a theory is the previously mentioned non minimally coupled scalar field theory \citep{Belgacem2021,Belgacem2022}.  In this case a natural ansatz for the correlations would be
\begin{equation}\label{eq:caseI}
    s_1(\|\vec{x}-\vec{y}\|) = \begin{cases}3 - 2 \left(\frac{r}{r_0}\right)^{n_{\text{DE}}}, \quad \quad r \leq r_0, \\ 1, \quad \quad \quad \quad \quad \quad \quad \ \  r > r_0.
    \end{cases} \quad \quad \quad \text{Case I.}
\end{equation}
Here, $r=\|\vec{x}-\vec{y}\|$ is the comoving distance between the coordinates. The factor 3 can be understood from Wick's theorem, as we assumed our field $\hat{\phi}$ to be Gaussian: if $\langle \hat{\rho}_Q(x) \hat{\rho}_Q(y)\rangle \propto \langle \hat{\phi}^2(x)\hat{\phi}^2(y)\rangle$ then the appropriate contractions give a factor 3 at coincidence. From Wick's theorem we expect that at $r\rightarrow\infty$ the function $s(\|\vec{x}-\vec{y}\|)$ approaches 1, as two out of three contractions contain correlators decay with distance. This is a consequence of the expectation that at $\|\vec{x}-\vec{y}\|\rightarrow\infty$ the fields decouple. In the intermediate regime we assume power law behaviour with slope\footnote{When calculating the 3D power spectrum associated to this correlation function, one sees that it is similar to the power spectrum from inflation where $n_{\rm DE}$ would be similar to the spectral index with a minus sign.} $n_{\operatorname{DE}}$ decaying with distance until at $r=r_0$ the fields are completely decoupled. Hereafter the correlations would saturate, only producing white noise.

Another possibility would be that the dark energy density field would be connected to a scalar field via $\hat{\rho}_Q\propto\hat{\phi}$. The density is then directly related to the vacuum expectation value of the field, which in this case we assume to be non zero. In this case Wick's theorem is of no help, as we do not have a 4-point correlator. We then assume that at coincidence the field $\hat{\phi}$ has a local variance $\sigma^2(z)=\langle\hat{\phi}^2\rangle(z)-\langle\hat{\phi}\rangle^2(z)$ and assume the fields decouple at large distances to the square of the vev. In the intermediate regime we again assume power law behaviour decaying with distance. For the energy density $\hat{\rho}_Q$ we then obtain the following $s$-functions:
\begin{equation}\label{eq:caseII}
    s_2(\|\vec{x}-\vec{y}\|) = \begin{cases}{\tilde{\sigma}_Q^2}
    \left[1 -  \left(\frac{r}{r_0}\right)^{n_{\text{DE}}}\right]+1, \quad \quad \quad \quad  r \leq r_0, \\ 1, \quad \quad \quad \quad \quad \quad \quad \quad \quad \quad \quad \quad \ \ \    r > r_0,
    \end{cases} \quad \quad \quad \text{Case II.}
\end{equation}
$r_0$ and $n_{\rm DE}$ are in principle different constants from \textit{Case I}. $\tilde{\sigma}^2_Q=\frac{B(z)^2\sigma^2}{\langle\rho_Q\rangle^2}$, with $B(z)$ the proportionality constant defined by $\hat{\rho}_Q=B(z)\hat{\phi}$. We keep the variance $\sigma^2$ as an independent variable. A specific theory could give a prediction for this variance. After $r>r_0$ we assume the fluctuations to saturate. We note two special cases: First, in the case the local variance goes to zero the classical limit is recovered. Second, when $\tilde{\sigma}_Q^2=2$ the $s-$functions of both cases are the same. The latter does not mean the theories are the same; in fact, the theories are still very different as in \textit{Case I} the density field is fundamentally non-Gaussian while in \textit{Case II} the density field is Gaussian. 

\begin{figure}
    \centering
    \includegraphics[width=0.7\textwidth]{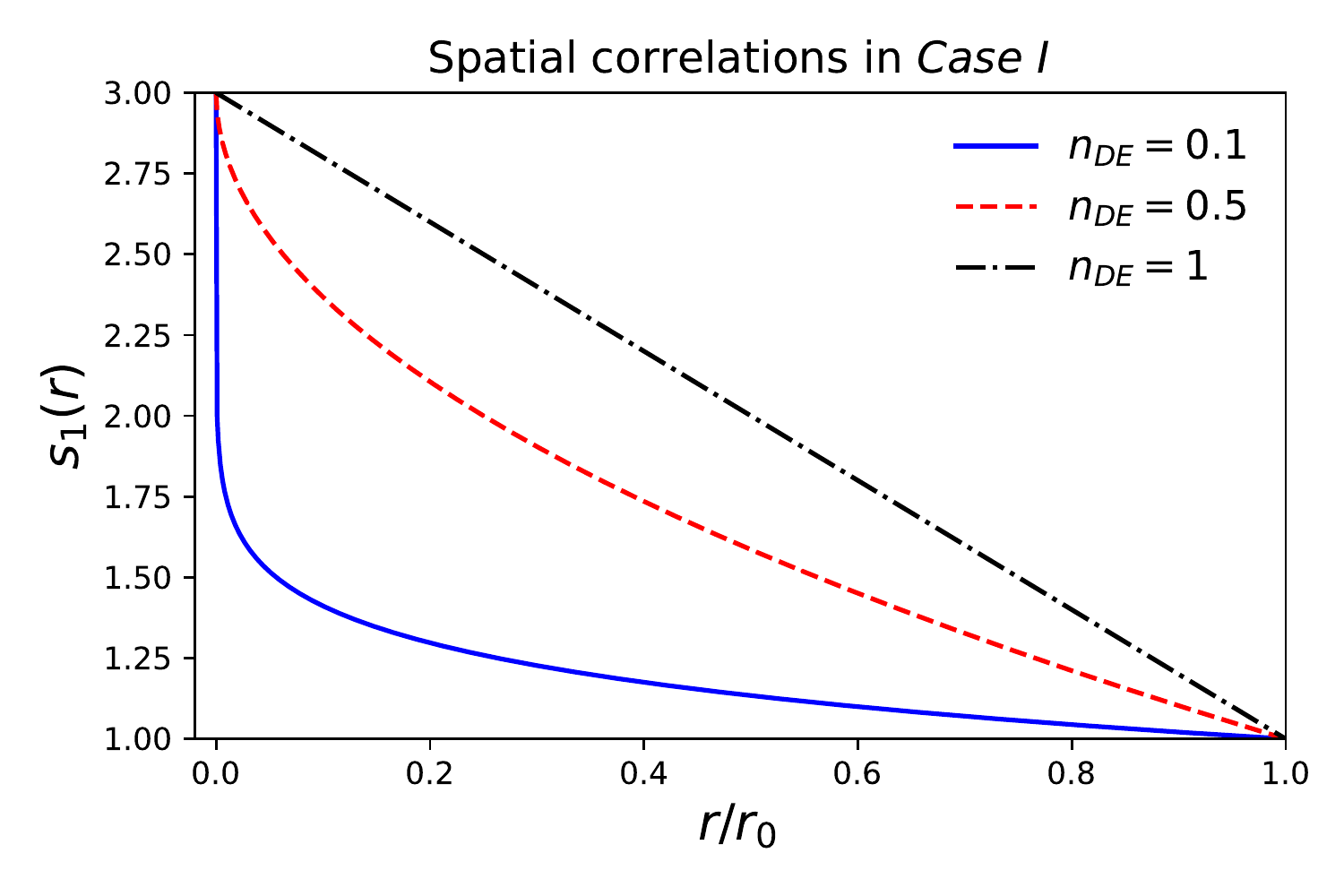}
    \caption{The function $s_1(r)$, describing the form of the spatial correlations in \textit{Case I}, for several values of the index $n_{\rm DE}$. The Figure gives $r$ in units of $r_0$. We expect $r_0$ to be of order $H_0^{-1}$ or larger.}
    \label{fig:sr}
\end{figure}

The function $s_1(r)$ for \textit{Case I} is shown in Figure \ref{fig:sr}. The function $s_2(r)$ has a similar shape, but goes from $1+\tilde{\sigma}^2_Q$ at coincidence to $1$ at $r=r_0$.
\section{Fluctuations in the luminosity distance}\label{sec:luminositydistance}

In this section we develop the equations for the luminosity distance in our setting including spatial correlations in dark energy. To see the effect of a fluctuating Hubble rate we first expand it in fluctuations $\hat{\rho}_Q = \langle \hat{\rho}_Q\rangle +\delta \hat{\rho}_Q$, where again the brackets denote an ensemble average, with this we obtain:
\begin{equation}\label{eq: friedmann}
    3 M_{P}^{2} \hat{H}^{2}(z, n)=\rho_{\text{tot}}(z) +\delta \hat{\rho}_Q(z,n) = \rho_{\text{tot}}(z)\left(1+\frac{\delta \hat{\rho}_Q(z,n)}{\rho_{\text{tot}}(z)} \right). 
\end{equation}
Where $\rho_{\text{tot}}(z)=\rho_c(z)+\langle\hat{\rho}_Q(z)\rangle$, is the total energy budget. Note that in this way taking the state average of this equation reduces it to the Friedmann equation in semi-classical gravity. In this treatment we thus go beyond semi-classical gravity by also including fluctuations on this. To be precise, this $\langle \hat{H}^2 \rangle$ is the square of the Hubble rate given by a state average defined over a spacelike hypersurface of constant time. Due to causality we can only observe over our past lightcone, which prohibits us from directly observing this $H$, as we cannot directly observe all of the structure on this hypersurface.

To connect with observations, we take the state average $\langle\ldots\rangle$ of the operator luminosity distance as defined by (\ref{eq:obsdis}): 
\begin{equation}\label{eq: luminositydistance}
    \frac{\langle \hat{d}_L\rangle (z)}{(1+z)} = \int_0^z d z^\prime \left\langle {\hat{H}(z^\prime,\hat{n})}^{-\frac{1}{2}}\right\rangle =  \int_0^z  \frac{dz^\prime}{\bar{H}(z^\prime)}\left \langle{\left({1+\frac{\delta \hat{\rho}_Q(z^\prime,n)}{\rho_{\text{tot}}(z^\prime)}}\right)^{-\frac{1}{2}}} \right \rangle.
\end{equation}
Here, $\sqrt{3M_P^2}\bar{H}=\sqrt{\rho_{\text{tot}}}$ is the root mean squared of (\ref{eq: friedmann}) and we also defined $\bar{H}(z)=\sqrt{\langle\hat{H}^2(z)\rangle}$. The second equality follows from substituting the square root of (\ref{eq: friedmann}) for $\hat{H}(z,\hat{n})$.  We can Taylor expand the term in brackets, as the ratio $\frac{\delta \hat{\rho}_Q(z,n)}{\rho_{\text{tot}}(z)}$ should be small. The linear term is per definition zero, thus we go to quadratic order in $\delta \hat{\rho}_Q$. This formal Taylor series diverges as a consequence of the Wick contractions of the higher n-point correlators going to infinity, but this does not give any problems as up to quadratic order it can still be trusted. In Appendix \ref{app:nonp} we derive via an alternative route the expansion for both cases up to all orders, from this we conclude that the error made by truncating at quadratic order is reasonable. This expansion gives us:
\begin{align}
    \frac{\langle \hat{d}_L\rangle (z)}{(1+z)} &=\int_0^z  \frac{dz^\prime}{\bar{H}(z^\prime)}\left(1 +\frac{3}{8}\frac{\langle \hat{\rho}_Q (z^\prime,\hat{n})^2 \rangle -\langle \hat{\rho}_Q (z^\prime) \rangle^2}{\hat{\rho}_{\text{tot}}(z^\prime)^2} \right)\\ &=\int_0^z  \frac{dz^\prime}{\bar{H}(z^\prime)}\left(1 + \frac{3}{8}\Omega_Q(z^\prime)^2\left(s_i(0)-1\right)\right).\label{eq:singlecorrelator}
\end{align}
Here, we defined $\Omega_Q(z)=\frac{\langle \rho_Q(z)\rangle}{\rho_\text{tot}(z)}$. The factor $s_i(0)-1$, where the subscript $i$ can take the values 1 or 2, referring to the case, can be interpreted as a normalised variance of the density field, as it is proportional to the variance of $\hat{\rho}_Q$. When $s_i(0)=1$ the local variance is zero and we recover the classical expression. This is the classical limit $\tilde{\sigma}^2_Q\rightarrow0$ of \textit{Case II}. However, in \textit{Case I} it is inevitable that the local variance is naturally large due to Wick's theorem and our density field being inherently non-Gaussian ($\hat{\rho}_Q\propto\hat{\phi}^2$). In both cases a deviation from the classical $s(0)=1$ results in an extra contribution to the luminosity distance, as the luminosity distance becomes dressed by the fluctuations. An interesting consequence of this is that the observed values do not necessarily match the global values of the underlying Friedmann-Lemaître-Robertson-Walker (FLRW) universe. We thus need to make the distinction between the local values and the global, or bare, quantities. Indeed, as we show in section \ref{sec:matching} for \textit{Case I} these quantities can be quite different as the dressed luminosity differs significantly from the classical one. In \textit{Case II} these values can also be different, however, the difference from the classical result is dictated by the parameter $\tilde{\sigma}_Q^2$, which in principle could be small.

Spatial correlations become visible in the two-point correlation function, defined as:
\begin{equation}\label{eq: luminositydistancecorrelator}
    \frac{\langle \hat{d}_L(z_1, \hat{n}_1) \hat{d}_L(z_2, \hat{n}_2) \rangle}{(1+z_1)(1+z_2)} = \int_0^{z_1} dz^{\prime} \int_0^{z_2} dz \left \langle {\hat{H}(z,n_1)^{-1} \hat{H}(z^\prime, n_2)^{-1}} \right \rangle.
\end{equation}
Following the same procedure of Taylor expanding and using the results in Appendix \ref{app: unequaltimecorrelator} we obtain:
\begin{align}
    \frac{\langle \hat{d}_L(z_1, \hat{n}_1) \hat{d}_L(z_2, \hat{n}_2)\rangle}{(1+z_1)(1+z_2)}  = & \int_0^{z_1} dz \int_0^{z_2} dz^\prime \frac{1}{\bar{H}(z)\bar{H}(z^\prime)} \left[1 +  \frac{3}{8}\left(\Omega_Q(z)^2 +\Omega_Q^2(z^\prime)\right)\left(s_i(0)-1\right) \right. \nonumber \\ &\left. +\frac{1}{4}\left(\frac{H_0^4}{\bar{H}(z)^2 \bar{H}(z^\prime)^2}\Omega_{Q,0}^2 \left(1+ (\Delta t(z) + \Delta t(z^\prime))\frac{\langle\dot{\hat{\rho}}_Q\rangle_0}{\langle\hat{\rho}_Q\rangle_0}\right)s_i(\|\vec{x}-\vec{y}\|) \right.\right.\nonumber \\&\left. \left.-\Omega_Q(z) \Omega_Q(z^\prime)\right) \right],\label{eq:finalld}
\end{align}
where $\Delta t(z) = t(z)-t_0$, $t(z)$ denotes the cosmological time at redshift $z$, the subscript 0 means it is evaluated today. $\dot{\hat{\rho}}_Q(t_0)$ denotes the time derivative of $\hat{\rho}_Q$ evaluated today. We can now see that the correlation function $s(\|\vec{x}-\vec{y}\|)$ has appeared, meaning this signal has angular dependence. When $s_i(0)\neq1$, as in \textit{Case I}, we can also identify another contribution representing the contribution of the local variance.

\section{Angular power spectrum}\label{sec:ps}
In this section we calculate the angular power spectrum related to the fluctuations of the luminosity distance. The angular power spectrum $C_\ell$ relates the correlations to a multipole expansion and provides a useful way to quantify the fluctuations in a way that is measurable \citep{Cooray_2010,Garoffolo2021}. We are mainly interested in the largest scales, as the dark energy correlations are strongest in this regime (which can be traced to the large characteristic length scale coming from inflation, see Section \ref{sec:spatcorr}). To study this regime consistently, we need a function that can be measured on the full sphere without approximations (\textit{e.g.} the flat sky approximation). The  $C_\ell$'s fulfill this need as we  are able to calculate them including wide angle effects. 

In our case the relevant field is the field corresponding to the luminosity distance fluctuations with respect to the mean. This is a scalar function and can be expanded in spherical harmonics $Y_{\ell m}$:
\begin{equation}
    \frac{\hat{d}_L(z,\hat{n})-\langle \hat{d}_L \rangle(z)}{\langle \hat{d}_L \rangle(z)} =\Delta(z,\hat{n})= \sum_{\ell=0}^\infty \sum_{m=-\ell}^\ell \Delta_{\ell m}(z) Y_{\ell m}(\hat{n}),
\end{equation}
This expansion defines the coefficients $\Delta_{\ell m}$. Assuming the process generating the fluctuations is statistically isotropic, the ensemble average of the $\Delta_{\ell m}$'s takes the following form:
\begin{equation}
        \langle \Delta_{\ell m}(z) \Delta^*_{\ell^\prime m^\prime}(z^\prime)\rangle = C_\ell(z,z^\prime) \delta_{\ell \ell^\prime}\delta_{m m^\prime},
\end{equation}
where $C_\ell(z,z^\prime)$ is the angular power spectrum. Because of statistical isotropy we can assume that independent $m$'s do not contain individual information and we can average over them. This gives us the following expression for the power spectrum:
\begin{equation}
    C_\ell(z_1,z_2)=\frac{1}{2\ell+1}\sum_{m=-\ell}^\ell\langle \Delta_{\ell m}(z_1)\Delta^*_{\ell m}(z_2)\rangle.
\end{equation}
 Due to the symmetries of the FLRW background the angular dependence only appears through the relative angle. Using the addition formula for spherical harmonics the $C_\ell$'s can be written as:
\begin{equation}
    C_\ell(z_1,z_2)= 2\pi \int_{-1}^1 d\mu \ \langle  \Delta(z_1) \Delta(z_2) \rangle(\mu) \mathcal{P}_\ell(\mu)\label{eq:cls}.
\end{equation}
Here, $\mu=\hat{n}\cdot\hat{n}^\prime=\cos \theta$, where $\theta$ is the relative angle between $\hat{n}$ and $\hat{n}^\prime$ and $\mathcal{P}_\ell(\mu)$ are the Legendre polynomials. Using the results from section \ref{sec:luminositydistance} we can write the $C_\ell$'s for $\ell \geq 1$ as\footnote{Here we assumed the regime where the fluctuations saturate is never reached. For our predictions later on this means we consider $H_0 r_0>2$.}:
\begin{align}
    C_\ell(z_1,z_2)=&-a_i \pi\frac{ \Omega_{Q,0}^2 H_0^4}{2}\frac{(1+z_1)(1+z_2)}{\langle \hat{d}_L \rangle (z_1) \langle \hat{d}_L \rangle (z_2)}\label{eq:clint} \\ & \times\int_{-1}^1 d\mu\int_0^{z_1} dz \int^{z_2}_0 dz^\prime \frac{\left(1+ (\Delta t(z) + \Delta t(z^\prime))\frac{\langle\hat{\dot{\rho}}_Q\rangle_0}{\langle\hat{\rho}_Q\rangle_0}\right)
    }{\bar{H}(z)^3 \bar{H}(z^\prime)^3} \left( \frac{r(z,z^\prime,\mu)^{n_{\rm DE}}}{r_0^{n_{\rm DE}}}\right)\mathcal{P}_\ell(\mu).\nonumber
\end{align}
 We introduced the constant $a_i$ to distinguish between the two cases, $a_1=2$ in \textit{Case I} and $a_2=\tilde{\sigma}_Q^2$ in \textit{Case II}. It is now possible to switch order of integration and integrate out the angular dependence $\mu$, using the results of Appendix \ref{app: hypergeometricintegral} we then obtain:
\begin{align}
    C_\ell(z_1,z_2) &=-\frac{a_i}{2}\pi^{\frac{3}{2}}\Omega_{Q,0}^2 H_0^4 \frac{2^{-\ell}\left(-\frac{{n_{\rm DE}}}{2}\right)_\ell}{\Gamma(\frac{3}{2}+\ell)}\frac{(1+z_1)(1+z_2)}{\langle \hat{d}_L \rangle (z_1) \langle \hat{d}_L \rangle (z_2)}\nonumber\\&\times \int_0^{z_1} dz \int_0^{z_2} dz^\prime\frac{\left(1+ (\Delta t(z) + \Delta t(z^\prime))\frac{\langle\hat{\dot{\rho}}_Q\rangle_0}{\langle\hat{\rho}_Q\rangle_0}\right)}{\bar{H}(z)^3 \bar{H}(z^\prime)^3}\left(\frac{\chi(z)^2+{\chi(z^\prime)}^2}{r_0^2}\right)^\frac{{n_{\rm DE}}}{2}\label{eq:clwrongnorm} \\&\times \mu_0(z,z^\prime)^{-\ell} {_2}F_1\left(\frac{\ell}{2}-\frac{{n_{\rm DE}}}{4},\frac{1}{2}+\frac{\ell}{2}-\frac{{n_{\rm DE}}}{4};\frac{3}{2}+\ell; \mu_0(z,z^\prime)^{-2}\right).\nonumber
\end{align}
Here, $\chi(z)=\int_0^z dz^\prime \bar{H}^{-1}(z^\prime)$ is the comoving coordinate. We used the short notation $\mu_0(z,z^\prime)=\frac{\chi(z)^2+\chi(z^\prime)^2}{2\chi(z)\chi(z^\prime)}$.  ${_2}F_1$ is Gauss' hypergeometric function and $(a)_\ell = \frac{\Gamma(a+\ell)}{\Gamma(a)}$ is the Pochammer symbol. This power spectrum cannot yet be compared with observations. Observations are done by measuring $\frac{\hat{d}_L(z,\hat{n})-  \bar{d_L} (z)}{ \bar{d}_L(z) }$, where
\begin{equation}\label{eq:angav}
    \bar{d}_L = \int \frac{d^2 \hat{n}}{4\pi} \hat{d}_L(z,\hat{n})
\end{equation}
is the luminosity distance averaged over all directions. If the angular averaging is the same as the ensemble averaging, the field is said to be ergodic. In our case, however, this is not the case. It is possible to relate our power spectrum to the one that would be observationally available. This can be approximated as a simple rescaling of our power spectrum and is derived in detail in Appendix \ref{app:ergodic}. The result is
\begin{align}
    \tilde{C}_\ell(z_1,z_2) =&-\frac{\frac{a_i}{2}}{1+\frac{C_0(z_1,z_2)}{4\pi}}\pi^{\frac{3}{2}}\Omega_{Q,0}^2 \bar{H}^4_0 \frac{2^{-\ell}\left(-\frac{{n_{\rm DE}}}{2}\right)_\ell}{\Gamma(\frac{3}{2}+\ell)}\frac{(1+z_1)(1+z_2)}{\langle \hat{d}_L \rangle (z_1) \langle \hat{d}_L \rangle (z_2)}\nonumber\\ & \times \int_0^{z_1} dz \int_0^{z_2} dz^\prime\frac{\left(1+ (\Delta t(z) + \Delta t(z^\prime))\frac{\langle\hat{\dot{\rho}}_Q\rangle_0}{\langle\hat{\rho}_Q\rangle_0}\right)}{\bar{H}(z)^3 \bar{H}(z^\prime)^3}\left(\frac{\chi(z)^2+{\chi(z^\prime)}^2}{r_0^2}\right)^\frac{{n_{\rm DE}}}{2}\label{eq:c(zz)} \\& \times \mu_0(z,z^\prime)^{-\ell} {_2}F_1\left(\frac{\ell}{2}-\frac{{n_{\rm DE}}}{4},\frac{1}{2}+\frac{\ell}{2}-\frac{{n_{\rm DE}}}{4};\frac{3}{2}+\ell; \mu_0(z,z^\prime)^{-2}\right),\nonumber
\end{align}
where $C_0(z,z^\prime)$ is the monopole of our angular power spectrum, which can be derived from (\ref{eq:cls}) and $\tilde{C}_\ell$ is the observable angular power spectrum. This effect can be shown to only change the result by a few percent, depending on the redshift the angular power spectrum is evaluated at. 

Our angular power spectrum is a function of $\ell,z$ and $z^\prime$. Observationally, however, the $C_l$'s are split into redshift bins and integrated over some redshift distribution that traces the field
\begin{equation}
    {C}^{i,j}_\ell = \int_0^\infty d z p^i(z) \int_0^\infty dz^\prime p^{\ j}(z^\prime) \tilde{C}_\ell(z,z^\prime).
\end{equation}
Here, $p^i(z)=\frac{1}{N^i}\frac{dN^i}{dz}$, where $N^i$ is the number of supernovae in the redshift bin, denotes the normalised supernovae number distribution as a function of redshift. The superscript $i$ denotes the redshift bin that is being used. After we switch order of integration  ($\int_0^\infty dz \int_0^z dz^\prime \rightarrow \int_0^\infty dz^\prime \int_{z^\prime}^\infty dz $), our $C_\ell$'s take the following form:
\begin{align}
    C^{i,j}_\ell =&-\frac{a_i}{2}\pi^{\frac{3}{2}}\Omega_{Q,0}^2 H_0^4 \frac{2^{-\ell}\left(-\frac{{n_{\rm DE}}}{2}\right)_\ell}{\Gamma(\frac{3}{2}+\ell)}\nonumber\\&\times\int_0^{\infty} dz \int_0^{\infty} dz^\prime\frac{W^{i,j}(z,z^\prime)}{\bar{H}(z)^3 \bar{H}(z^\prime)^3}\left(1+ (\Delta t(z) + \Delta t(z^\prime))\frac{\langle\hat{\dot{\rho}}_Q\rangle_0}{\langle\hat{\rho}_Q\rangle_0}\right)\left(\frac{\chi(z)^2+{\chi(z^\prime)}^2}{r_0^2}\right)^\frac{{n_{\rm DE}}}{2}\nonumber\\&\times  \mu_0(z,z^\prime)^{-\ell} {_2}F_1\left(\frac{\ell}{2}-\frac{{n_{\rm DE}}}{4},\frac{1}{2}+\frac{\ell}{2}-\frac{{n_{\rm DE}}}{4};\frac{3}{2}+\ell; \mu_0(z,z^\prime)^{-2}\right)\label{eq:finalcl},
\end{align}
where $W^{i,j}(z,z^\prime)$ is defined as:
\begin{equation}
    W^{i,j}(z,z^\prime) =\int_z^\infty dz_1 \int^\infty_{z^\prime} dz_2  \frac{(1+z_1)(1+z_2)p^i(z_1)p^{\ j}(z_2)}{\langle \hat{d}_L \rangle (z_1) \langle \hat{d}_L \rangle (z_2)\left(1+\frac{C_0(z_1,z_2)}{4\pi}\right)}.
\end{equation}
\section{Connecting with observations}\label{sec:obs}

In this section we present the tools needed to connect this to observations. We relate the quantities in our model, which are bare quantities related to the underlying non perturbed FLRW universe, to measured quantities. We also consider other perturbation effects that would be contributing to this power spectrum, which we refer to as contaminant effects. These effects arise due to the interplay between the observed matter perturbations and general relativity. We then present the supernovae distributions used to make forecasts and describe the formalism used for the expected noise of these power spectra, which is crucial to assess its measurability.

\subsection{Connecting the bare parameters to observed ones}\label{sec:matching}
In the previously discussed \textit{Case I}, we obtained luminosity distance predictions which differ from the standard $\Lambda$CDM. Because of this the Hubble parameter and the density fraction measured locally differs from the one that would be measured on a spacelike hypersurface of equal time. Due to causality we can only perform measurements along our past light cone, thus we cannot access this global Hubble parameter and density fraction directly.

To estimate these bare parameters one would ideally fit our model to current data sets. For example, comparing it with current supernovae samples. We proceed in a similar fashion, we obtain our values by comparing it to the local low redshift distance ladder measurements from the \textit{SH0ES} collaboration \cite{Riess_2016}. This gives the following constraint up to cubic order:
\begin{equation}\label{eq:lowzexp}
    \langle \hat{d}_{L} \rangle(z)=\frac{ z}{H_{L}}\left\{1+\frac{1}{2}\left[1-q_{L}\right] z-\frac{1}{6}\left[1-q_{L}-3 q_{L}^{2}+j_{L}\right] z^{2}\right\}.
\end{equation}
Here, $q_L$ is known as the deceleration parameter and $j_L$ the jerk parameter, the subscript reminds that the values are locally measured. This expansion is model independent and therefore well suited for our approach. This can then be matched power by power. To this end we need to expand our luminosity distance (\ref{eq:singlecorrelator}) as well, including the extra term due to the included fluctuations. We assume that the background is well captured by flat $w$CDM and that the classical part of our model scales like non-relativistic matter as it contains the contributions from cold dark matter and baryons. Due to this parametrization our luminosity distance and energy density fraction take the following form:
\begin{align}
    \langle \hat{d}_L \rangle(z) &= \frac{1+z}{H_0}\int_0^z dz^\prime \frac{1+\frac{3}{4}\Omega_Q(z)^2}{\sqrt{\Omega_{c,0} (1+z)^3 +\Omega_{Q,0} (1+z)^{3(w_Q+1)}}},\\ \Omega_Q(z) &= \frac{\Omega_{Q,0}(1+z)^{3(w_Q+1)}}{\Omega_{c,0} (1+z)^3 +\Omega_{Q,0} (1+z)^{3(w_Q+1)}}.
\end{align}
Here $\Omega_{c,0}=1-\Omega_{Q,0}$. Expanding this and matching it power by power with (\ref{eq:lowzexp}) then gives the following three equations:
\begin{align}
\frac{1}{H_L}  = & \frac{1}{H_0}\left(1+\frac{3}{4}\Omega_{Q,0}^2\right),  \\
\frac{1}{2 H_L}\left[1-q_L\right]  = &\frac{1}{4 H_0}\left(1+\frac{3}{4}\Omega_{Q,0}^2-\frac{3}{4}w_Q\left(4-12\Omega_{Q,0}+15\Omega_{Q,0}^2\right)\right),  \\
-\frac{1}{6 H_L}\left[1-q_L-3 q_L^2 + j_L\right]  = &-\frac{1}{8H_0}\left(1+\frac{3}{4}\Omega_{Q,0}^2-\frac{w_Q}{2}\left(4-12\Omega_{Q,0}+15\Omega_{Q,0}^2\right)\right.\nonumber \\ & +\left.\frac{3w_Q^2}{4}\Omega_{Q,0}\left(8-60\Omega_{Q,0}+150\Omega_{Q,0}^2-105\Omega_{Q,0}^3\right)\right).
\end{align}
Once the local parameters are fixed by measurements, we can numerically solve this system of equations for $\Omega_{Q,0},H_0$ and $w_{Q}$.  For the local parameters we use the values found by \citep{Riess_2016}, these are obtained by fitting the low redshift expansion over the redshift range spanning from $z=0.023$ to $z=0.15$. They find that $H_{L}=74.1 \pm 1.3 \mathrm{~km} \mathrm{~s}^{-1} \mathrm{Mpc}^{-1},q_L=-0.55$ and $j_L=1$. The solutions for the bare parameters are then $H_0 = 118.0 \mathrm{~km} \mathrm{~s}^{-1} \mathrm{Mpc}^{-1}$, $\Omega_{Q,0}=0.89$ and $w_Q=-0.97$. We stress that this is not an actual fit, but an estimate to obtain predictions for the measurability in upcoming datasets.

\subsection{Contaminants from perturbed $\Lambda$CDM}\label{sec:contaminants}
To know whether or not our signal is measurable, it is crucial to compare it with the signal that would be observed in a null experiment, i.e. a universe without spatial correlations in dark energy. In this section we calculate the signal that would be measured in a such an universe. Consider a perturbed $\Lambda$CDM universe. 
Fluctuations in the luminosity distance have been well studied in this setting \cite{sasaki1987,Bonvin_2006,Biern_2017,Garoffolo2021}. These fluctuations are intrinsically relativistic effects known as convergence, Doppler, (integrated) Sachs-Wolfe, volume dilation and time delay \cite{sasaki1987,Bonvin_2006,Biern_2017,Garoffolo2021,Yoo_2009,Yoo_2010,Bonvin_2014}. These effects affect the magnitude of the supernova observed and thus alter the luminosity distance observed. These effects depend on the large scale structure and thus also lead to a non zero angular power spectrum.
It has been shown that all effects are subdominant to the convergence and Doppler effects \cite{Bonvin_2006}. Therefore we focus only on these effects. Both of these contributions are well studied, albeit mostly in the context of galaxy surveys \cite{Yoo_2009,Yoo_2010,Bacon_2014,Bonvin_2008,Bonvin_2014}. The effect in a supernova survey is in a sense more direct, as the observable is directly the magnitude of the source. Both Doppler and convergence are called lensing effects, as they affect the magnitude of the source.

Schematically we can write the Doppler and convergence contributions as
\begin{equation}\label{eq:dllcdm}
    \frac{{d}_L(z,\hat{n})}{\Bar{d}_L(z)} -1= \kappa_c + \kappa_v.
\end{equation}
These contributions are then given in standard $\Lambda$CDM. Here, $\bar{d}_L$ is the angular averaged luminosity distance (\ref{eq:angav}). 

The convergence contribution is given by \cite{sasaki1987}:
\begin{equation}
    \kappa_c(z,\hat{n})=\int_0^z \frac{dz^\prime}{H(z^\prime)}\frac{\chi(z)-\chi(z^\prime)}{\chi(z)\chi(z^\prime)}\Delta_\perp\left(\Phi(z,\vec{k})+\Psi(z,\vec{k})\right).
\end{equation}
Here $\Delta_\perp$ is the Laplacian evaluated transverse to  the line of sight and $\Phi$ and $\Psi$ are the Bardeen potentials. The convergence term expresses how certain overdense regions between the observer and the supernova magnify the source, thereby increasing (or decreasing) the supernova's luminosity. We note that this effect depends on the full line of sight and thus becomes stronger with redshift. 

The second term in (\ref{eq:dllcdm}) is known as the Doppler term, it is given by \cite{Bonvin_2006}:
\begin{equation}\label{eq:dopp}
    \kappa_v(z,\hat{n})=\left(\frac{1}{\chi(z)\mathcal{H}(z)}-1\right)\vec{v}\cdot\hat{n}.
\end{equation}
Here $\mathcal{H}(z)$ is the conformal Hubble parameter $a(z)H(z)$. The Doppler term reflects the effect given by the peculiar velocities of supernovae. The supernova is moving with respect to the observer, as a result of this the supernova appears (de)magnified. This effect depends on the velocity's direction, $\vec{v}\cdot\hat{n}$, reflecting the fact that the effect is opposite if it is either moving from or towards the observer.  At low redshift the first term in (\ref{eq:dopp}) dominates, due to the small comoving distance $\chi$. If we assume the supernova is moving towards the observer, meaning $\vec{v}\cdot\hat{n}<0$. Then the Doppler term demagnifies the supernova at low redshift. At high redshift the second term dominates and as a result the supernova is magnified in this regime. This can be explained as follows \citep{Bonvin_2014}: when fixing the redshift, a supernova moving towards us is in reality more distant in comoving coordinates than it appears. At low redshift this gives a negative contribution, due to the smaller angle it is observed under. At high redshift, however, this is not the case. In this regime the dominant contribution comes from the universe having a smaller scale factor when the photons were emitted. The bundle of photons is then stretched under the expansion of the universe when moving towards us, increasing the observed magnitude. For a \textit{Planck} cosmology \citep{planck2018} the effects are equal at $z\approx1.6$ and cancel out. For the values of $z$ that we consider, the low redshift contribution always dominates. This effect then becomes smaller with redshift. 

To connect the velocity field to the matter density field $\delta(\vec{k},z)$, which in turn can be connected to the power spectrum, we need the continuity equation \cite{dodelson2020modern}: 
\begin{equation}\label{eq:cont}
    \Vec{v}(\vec{k},z) = i \mathcal{H}(z)f(z)\frac{\Vec{k}}{k^2}\delta(\Vec{k},z).
\end{equation}
Here, $f(z)= \frac{d \ln D(z)}{d \ln a}$ is the growth rate with $D(z)$ being the growth function.

To compare with our main effect we calculate the angular power spectra $C_\ell^{v}$ and $C_\ell^{c}$, the power spectra coming from the auto correlation functions of both effects. We expect the cross-correlation between the Doppler and convergence effect to be zero, as the Doppler effect selects modes along the line of sight, while the convergence effect selects modes perpendicular to the line of sight. The power spectra are derived in Appendix \ref{app:dopplerlensing} \cite{sasaki1987,Bonvin_2006,Bonvin_2014,Bacon_2014}. For the Doppler effect the spectrum is given by:
\begin{multline}
    C^{v,i,j}_\ell=  \frac{2}{\pi}\int_0^\infty dz_1 p^i (z_1) \int_0^\infty dz_2 p^{\ j} (z_2)\left(\frac{1}{\chi(z_1)}-\mathcal{H}(z_1)\right)f(z_1)\left(\frac{1}{\chi(z_2)}-\mathcal{H}(z_2)\right)f(z_2) \\ \times\int_0^\infty dk P_m(k,z_1,z_2) j_{\ell}^{\prime}(k \chi(z_1))  j_{\ell}^{\prime}(k \chi(z_2)).
\end{multline}
While for the convergence spectrum the result takes the following form:
\begin{multline}\label{eq:lensingfinal}
    C^{c,i,j}_\ell = \frac{2}{\pi}(\ell(\ell+1))^2 \int_0^\infty  \frac{dz_1}{H(z_1)}W^{\ i}_L(z_1)\int_0^{\infty} \frac{dz_2}{H(z_2)}W^{\ j}_L(z_2)\\ \times \int_0^\infty dk k^{2} T_{\Phi+\Psi}(k,z_1) T_{\Phi+\Psi}(k,z_2) P_m(k,z_1,z_2) j_\ell(k\chi(z_1)) j_\ell(k\chi(z_2)). 
\end{multline}
With $W^i_L(z_i)$ being the lensing kernel defined as:
\begin{equation}
    W^i_L(z_i) = \int_{z_i}^\infty dz p^i (z) \frac{\chi(z)-\chi(z_i)}{\chi(z)\chi(z_i)}.
\end{equation}
 $i$ and $j$ denote the redshift distribution the field is sampled over. $P_m(k,z_1,z_2)$ is the matter power spectrum as a function of wavenumber $k$ and redshift $z_1$ and $z_2$ and is defined by $\langle\delta(\vec{k},z)\delta^*(\vec{k}^\prime,z^\prime)\rangle = (2\pi)^3 P_m(k,z,z^\prime)\delta^{3}(\vec{k}-\vec{k^\prime})$.  $T_{\Phi+\Psi}$ is the transfer function relating the dark matter field to the Bardeen potentials. To simplify the computations we use the linear power spectrum at the largest scales instead of the non linear spectrum. After $\ell=15$ we switch to the non linear power spectrum. To still efficiently calculate these integrals we then adopt the Limber approximation \citep{limber1953}. Switching at $\ell=15$ ensures an almost smooth transition ($ < 1\%$) between the calculation methods. The power spectrum, growth rate, growth function and transfer functions are obtain by using the publicly available Core Cosmology Library ({\tt CCL} \cite{Chisari_2019}, v2.1.0), {\tt CCL} uses {\tt CAMB} \cite{camb} to obtain predictions for the power spectra.
 
 For these quantities we assume the following flat \textit{Planck} $\Lambda$CDM cosmology \citep{planck2018}: $\Omega_b=0.045$, $\Omega_{\rm CDM} = 0.27$, $h = H_0/(100\,{\rm km\,s^{-1} Mpc^{-1}})=0.67$, $\mathcal{A}_s=2.1\times10^{-9}$ on a pivot scale of $k_p=0.05 \ {\rm Mpc^{-1}}$ and $n_s = 0.96$. In $\Lambda$CDM $\Omega_b$ and $\Omega_{\rm CDM}$ are the fractional energy densities of baryonic and cold dark matter, $h$ is the dimensionless Hubble constant, $\mathcal{A}_s$ is the variance of curvature perturbations in a logarithmic wavenumber interval centered around the pivot scale $k_p$, and $n_s$ is the scalar tilt. 

Lastly, we note that in principle we could also have correlations coming from surveys being magnitude limited. This induces a bias known as the Malmquist bias, which affects the distribution $p(z)$. The lensing effect magnifies or demagnifies supernovae and can thus push stars that are otherwise to faint to observe over the magnification threshold. This effect depends on the large scale structure and thus induces non zero correlations. Specifically the observed number of supernovae $N$ in a redshift shell $z$ would be altered by $\frac{N(z,\hat{n})}{\bar{N}(z)}=1+5 s (\kappa_c+\kappa_v)$ , where $s$ is the effective number count slope  \citep{Broadhurst_1995,Moessner_1998}. This would then be a correction to the distribution $p^i(z)\rightarrow p^i(z)(1+5 s (\kappa_c+\kappa_v))$. However, this correction is higher order as it multiplies (\ref{eq:dllcdm}), which is already leading order in perturbation theory.
\subsection{Forecasts}
We aim to predict the overall detectability of the signal. To this end we provide a forecast for the signal to noise ratio (SNR). The SNR of an angular power spectrum $C_l$ between two redshift bins $i$ and $j$ is defined as:
\begin{equation}
   {\rm SNR}^{i j}=\sqrt{\sum_{l=l_{\min }}^{l=l_{\max }} \frac{\left(C^{ij}_{l}\right)^{2}}{\operatorname{Var}\left[C^{ij}_{l}\right]}}.
\end{equation}
Here, $\operatorname{Var}\left[C_{l}\right]$ is the variance of the power spectrum. Assuming all the perturbations are statistically homogeneous, isotropic and Gaussian the variance of a power spectrum is given by \citep{Asorey_2012}:
\begin{equation}\label{eq:noise}
    \operatorname{Var}\left[C_{l}^{i j}\right]=\frac{\tilde{C}_{l}^{ii} \tilde{C}_{l}^{jj}+\tilde{C}_{l}^{i j} \tilde{C}_{l}^{j i}}{(2 l+1) f_{\mathrm{sky}}},
\end{equation}
where $\tilde{C}^{ij}_\ell$ is the angular power spectrum between two redshift bins including noise and $f_{\rm sky}$ is the observed fraction of the sky. We consider noise for the auto-power spectra and we assume this to be white noise. This noise is a result from the fact that we do not measure a smooth field, but a finite number of supernovae. We refer to this noise as shot noise. For our power spectrum it takes the following form \citep{Garoffolo2021}:
\begin{equation}
    \tilde{C}^{ii}_\ell = C_\ell^{ii}+\frac{4\pi f_{\rm sky}}{N_{SNe}}\left(\frac{\sigma_{d_L}}{d_L}\right)^2.
\end{equation}
 $N_{SNe}$ is the amount of supernovae in the redshift bin and $\frac{\sigma_{d_L}}{d_L}$ is the intrinsic dispersion of luminosity distance measurement, this can be related to the intrinsic magnitude dispersion of the supernova via $\frac{\sigma_{d_L}}{d_L}=\frac{\ln 10}{5}\sigma_m$. The intrinsic uncertainty in the magnitude is usually estimated to be roughly 0.1-0.2 \cite{Scolnic2018,Amanullah_2003,des2012}. We assume an $\ell$ range of $2\leq \ell \leq 100$. However, we explore the dependence of the SNR on $\ell_{\rm min}$ later as well.
\subsection{Supernova sample}
\begin{figure}
    \centering
    \includegraphics[width=0.7\textwidth]{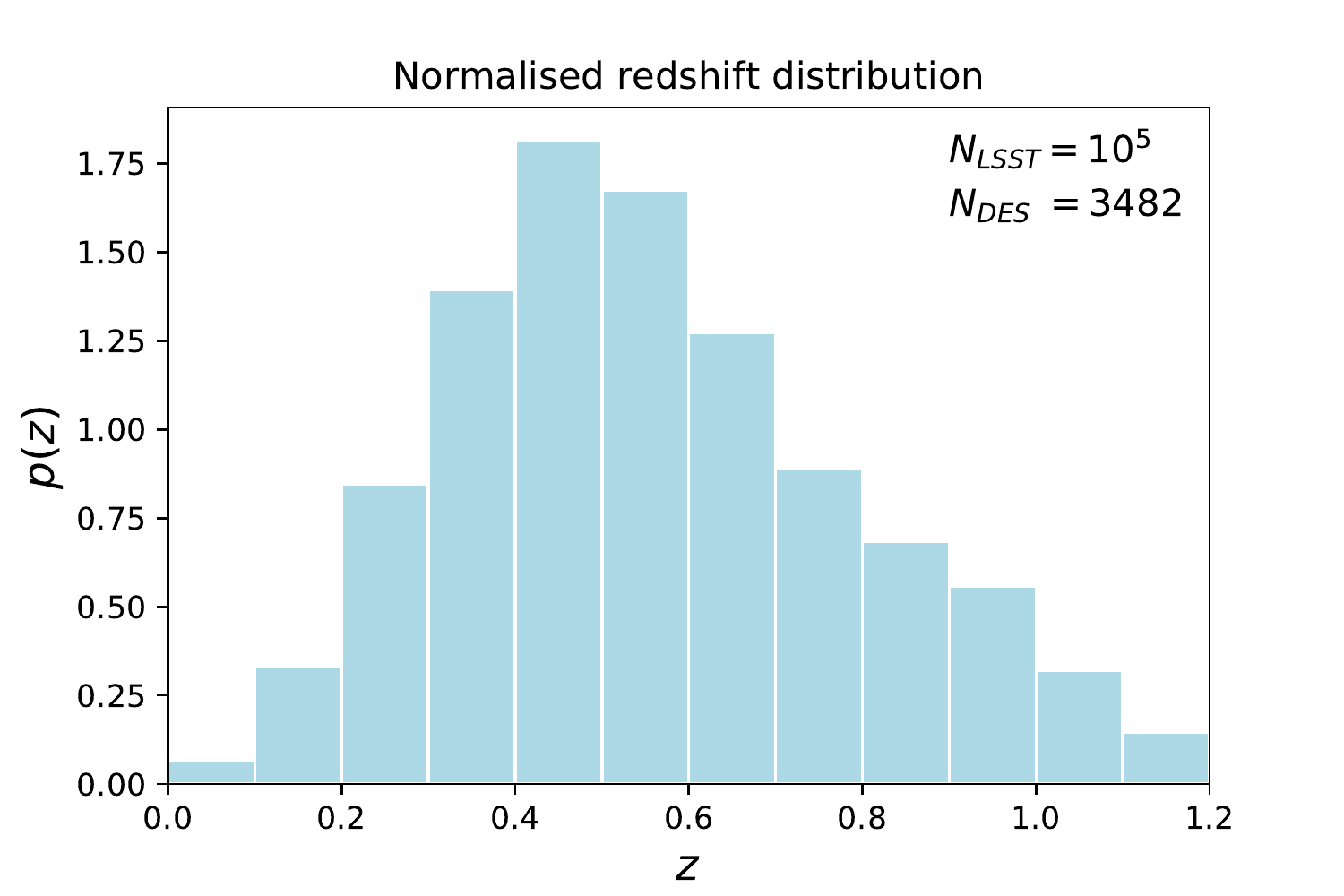}
    \caption{The supernova distribution expected to be obtained by the Dark Energy Survey (DES) \citep{des2012}. This survey will observe 30 square degrees of the sky repeatedly, 6 of which will be allocated more observing time for so called `deep drilling'. This survey will be used to make forecasts for the observability of our dark energy signal.}
    \label{fig:dessample}
\end{figure}

For the predictions of the $C_\ell$'s and the noise we need estimates of $f_{\rm sky}, p(z)$, $N_{SN_e}$ and $\sigma_{d_L}$. We obtain predictions for two different surveys: The ongoing Dark Energy Survey (DES) and the upcoming Legacy Survey of Space and Time (LSST). For DES we use the simulated supernova distribution as described by ref. \citep{des2012}, which makes predictions for the year 5 sample DES will produce for different strategies. The relevant strategy that produces the final year 5 sample is the Hybrid-10 strategy \citep{Vincenzi_2021}. In this case the telescope will visit 10 regions of 3 $\operatorname{deg}^2$, covering a total area of 30 $\operatorname{deg}^2$, corresponding with a fraction of the sky $f_{\rm sky}=0.0007$. Two of these regions will have a longer exposure time, performing `deep drills' into redshift space. This strategy is expected to find $\approx$ 3482 supernovae with accurate redshifts. This sample is shown in Figure \ref{fig:dessample}.

Separately we also make forecasts in a more futuristic setup, with the goal of mimicking the year 10 (Y10) LSST data. This sample will improve significantly on current samples for several reasons: it will have a significantly higher number count, with an expected number of supernovae with well measured redshift of about $10^5$ \citep{lsstscience}, almost two orders of magnitudes more than DES. Secondly, LSST will observe a much larger region of the sky, as LSST will produce the first all sky supernova survey, with $f_{\rm sky}=0.5$ \citep{lsstscience}. This will significantly reduce the cosmic variance. We assume the total LSST sample to be similar in shape as the DES sample, therefore we use the same redshift distribution $p(z)$. 

For both surveys we estimate the intrinsic uncertainty in magnitude $\sigma_m$ as 0.13, in accordance with \citep{des2012}.  
\section{Results}\label{sec:results}
In this section we present the results for the angular power spectra. We analyse which range of the model parameters $H_0 r_0$ and $n_{\rm DE}$  would lead to a possible positive detection, and assess what is the best strategy to detect this signal. We consider $H_0 r_0$ instead of $r_0$ as this is a dimensionless quantity, the length scale of the fluctuations with respect to the Hubble length today. We only consider lengths $r_0$ larger than twice the Hubble length, thus we never reach the regime where the fluctuations saturate. The general shape of the angular power spectrum does not significantly change as the factor $\left(H_0 r_0\right)^{-n_{\rm DE}}$ mostly affects the general amplitude, therefore not changing the form of the spectrum.

Our predictions are be made in the context of \textit{Case I}. Generalising the results to \textit{Case II} is in principle straightforward once $\tilde{\sigma}^2_Q$ is fixed. A simple comparison can be made in the following cases, when $\tilde{\sigma}^2_Q=2$ the results are the same and when $\tilde{\sigma}^2_Q<2$ ($\tilde{\sigma}^2_Q>2$)  \textit{Case II} would be harder (easier) to detect than the results in this section.
\begin{figure}
    \centering
    \subfloat[]{\includegraphics[width=0.5\textwidth]{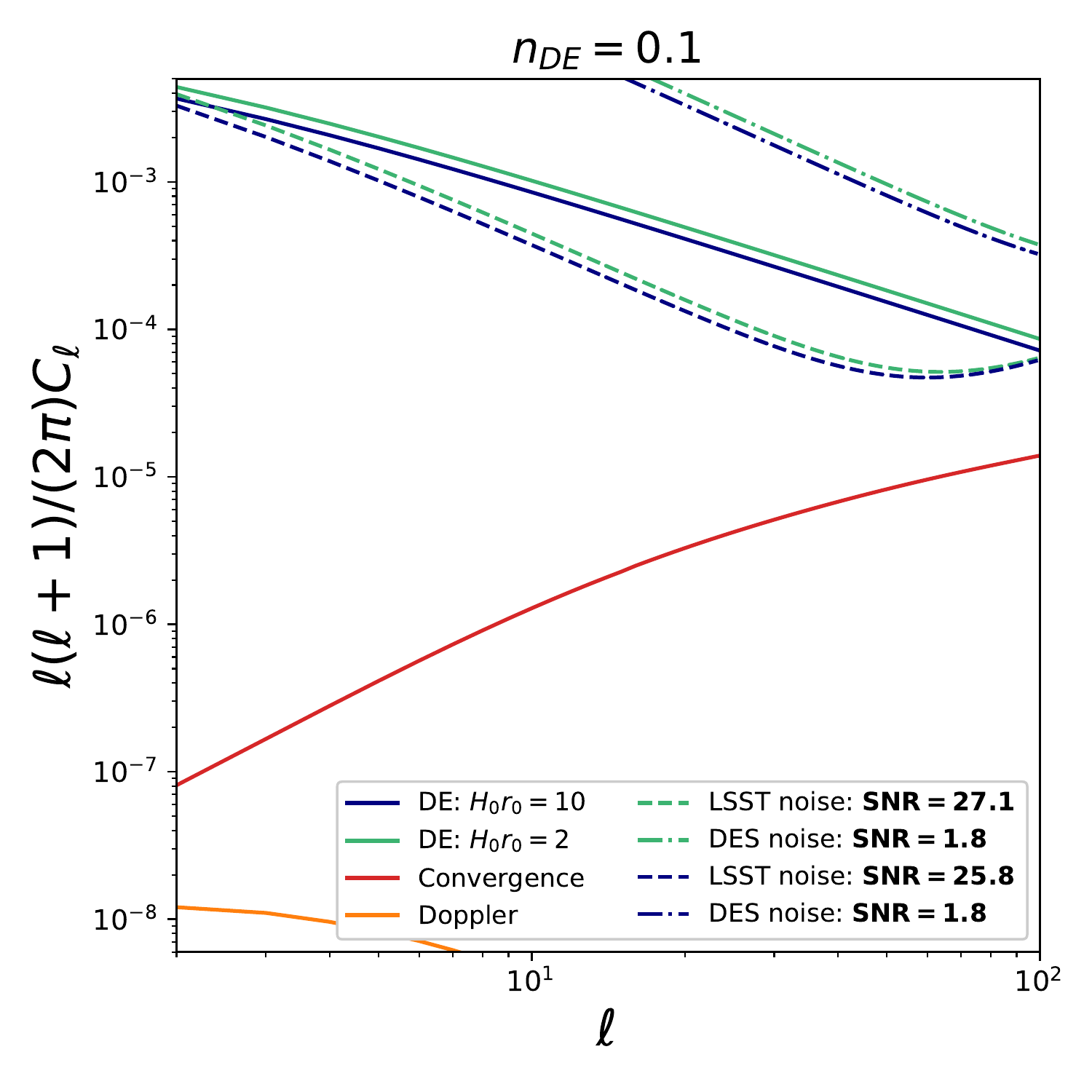}\label{fig:nde1}}
    \subfloat[]{\includegraphics[width=0.5\textwidth]{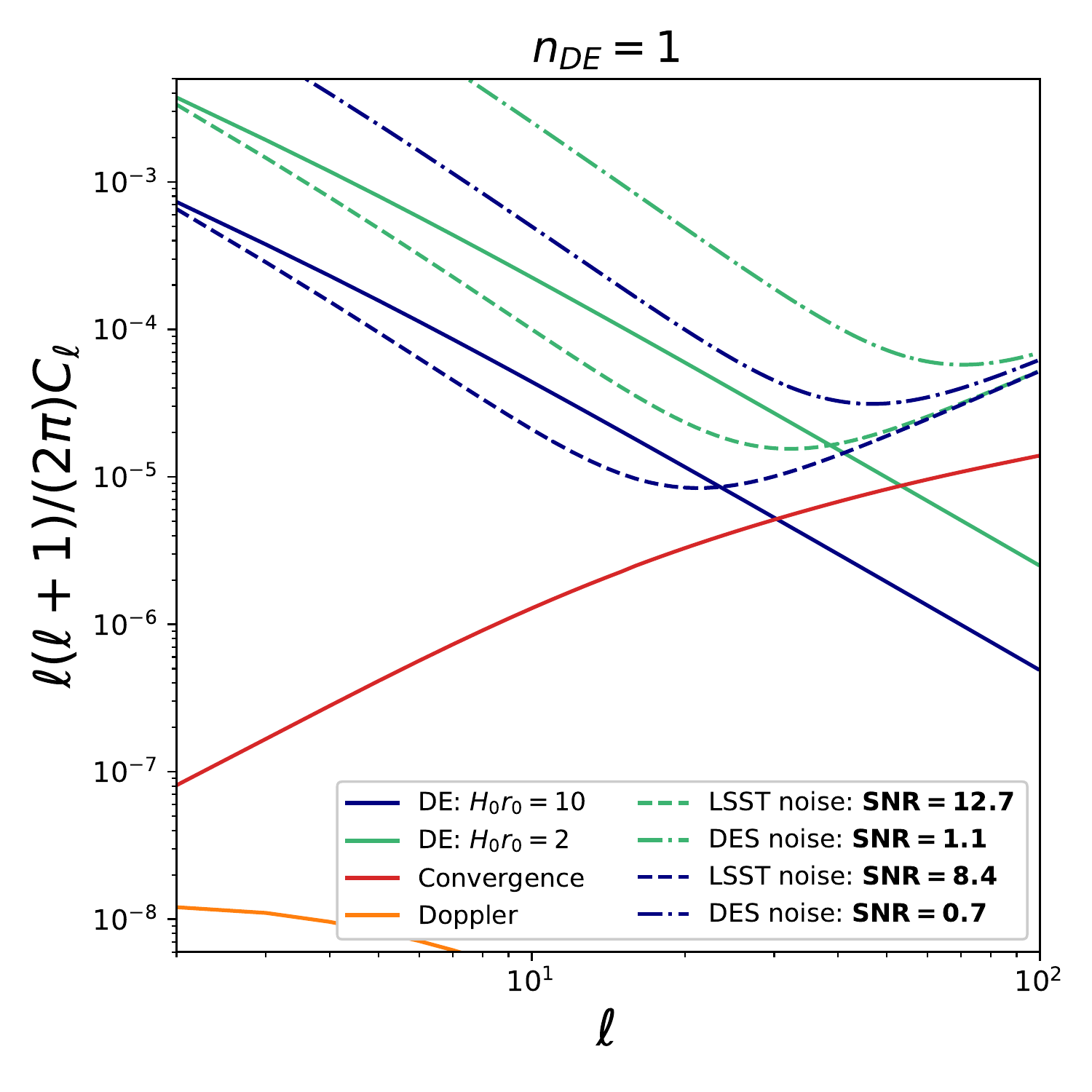}\label{fig:nde2}}
    \caption{The angular power spectra $C_\ell$ shown with characteristic fluctuation scale $H_0 r_0=2$ (solid green line) and $10$ (solid blue line). We also show the spectra for the lensing effects (convergence in red and Doppler in orange). The left panel shows the dark energy signal with a spectral slope of $n_{\rm DE}=0.1$ and the right panel the spectra for $n_{\rm DE}=1$. The figures also include the expected noise, including shot noise and cosmic variance for the two survey set ups (LSST Y10 - dashed and DES - dot-dashed). This gives four curves, as the noise depends both on the survey and the power spectrum via (\ref{eq:noise}). In the legend the expected signal to noise ratio SNR corresponding to this noise is given in bold for DES and LSST Y10.}
    \label{fig:nde}
\end{figure} 
\subsection{Overall detectability}
First, we assess the detectability over the full redshift distribution. In Figure \ref{fig:nde} the angular power spectrum (\ref{eq:finalcl}) is shown. It is clear that the signal is mostly located at the lower $\ell$'s and then decays very rapidly. This decay is steepest in the case where spectral index is larger,  which is the case shown in Figure \ref{fig:nde2}. We understand the fact that a higher $n_{\rm DE}$ gives a smaller signal as follows: due to always considering large scales, indicated by $H_0 r_0 \geq 2$, the fluctuations we are probing are those with a relatively small distance with respect to the length scale $r_0$. When we consider the profiles of the correlations, Figure \ref{fig:sr}, the profiles with a small spectral index change significantly more rapidly at smaller $r$. This then results in a higher $C_\ell$.

In the low $\ell$ regime, the main contribution of noise is the cosmic variance. The cosmic variance is especially large for DES due to the small fraction of the sky it includes (only 30 ${\rm deg^2}$). LSST does not have this problem as it includes all of the southern hemisphere for the Y10 survey. When comparing the noise for the power spectra (dashed (LSST)/dot-dashed (DES)) with the green and blue lines) we see that the prospects of detecting this effect with the dark energy survey (DES) are very small, as the low $\ell$ regime is drowned in cosmic variance. Even though the total amount of supernovae observed is significantly larger for LSST, the number density is actually quite similar to DES. The shot noise is thus comparable for both surveys. 
For LSST the signal rises above the noise for $\ell_{\rm max}=20$ when $n_{\rm DE}\approx1$ and $\ell_{\rm max}\approx100$ when $n_{\rm DE}=1$. 

In this regime we also observe that both the convergence as the Doppler effect are subdominant to the dark energy spectrum and is thus not a nuisance. In the case $n_{\rm DE}=1$, around $\ell\approx40$ the convergence effect does surpass the dark energy signal; however, in this regime the signal is located below the shot noise, so it would not contribute to a detection. The Doppler signal is similar to the dark energy signal as it is mainly relevant at large scales. However, even at these scales it is several orders of magnitude below the dark energy signal and thus can safely be neglected. We also note that in general the lensing effects are below the shot noise, thus in the regime where our signal would be of the same order as the lensing effects (large $H_0 r_0$) the signal is not measurable in any case.

\begin{figure}
    \centering
    \subfloat[]{\includegraphics[width=0.5\textwidth]{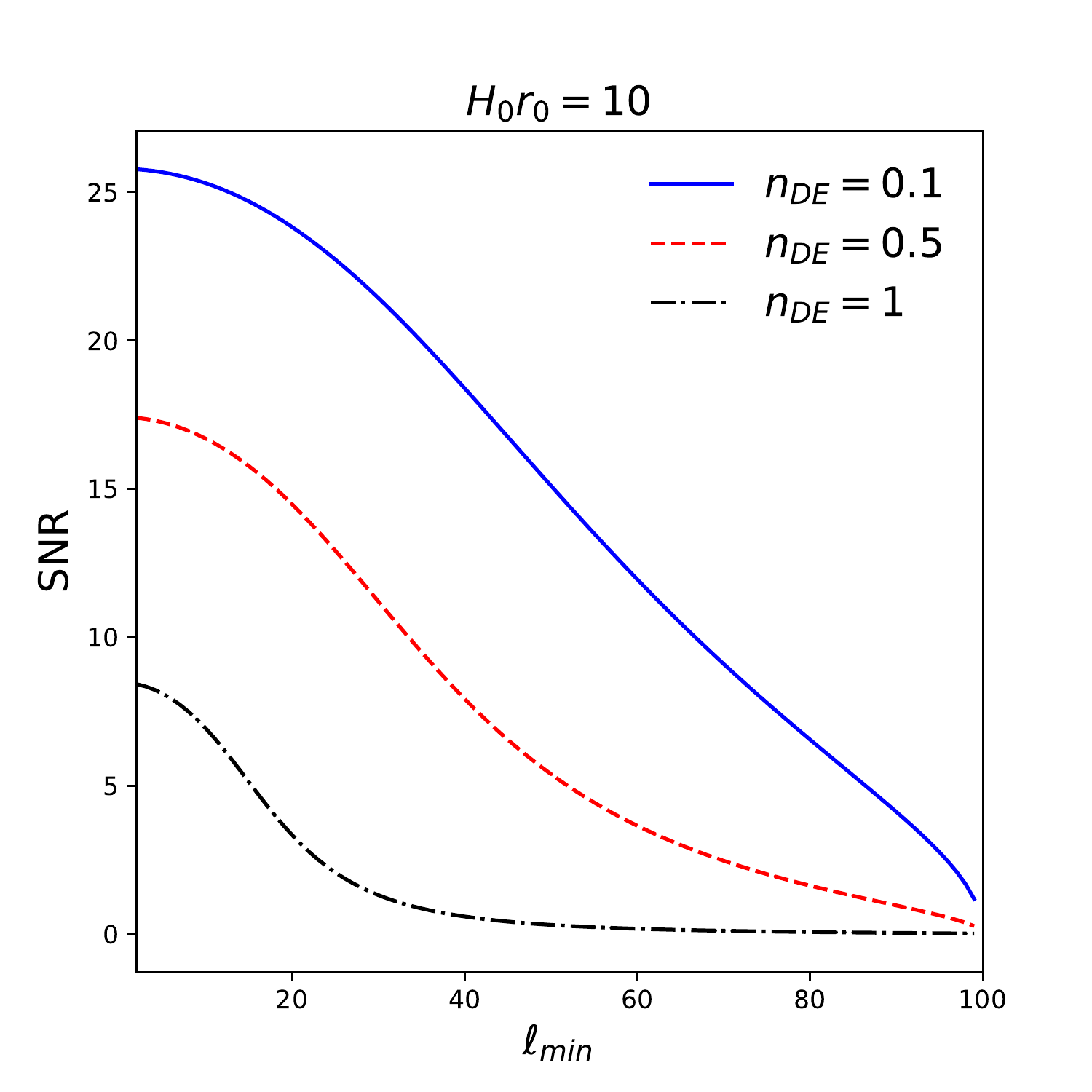}\label{fig:sn1}}
    \subfloat[]{\includegraphics[width=0.5\textwidth]{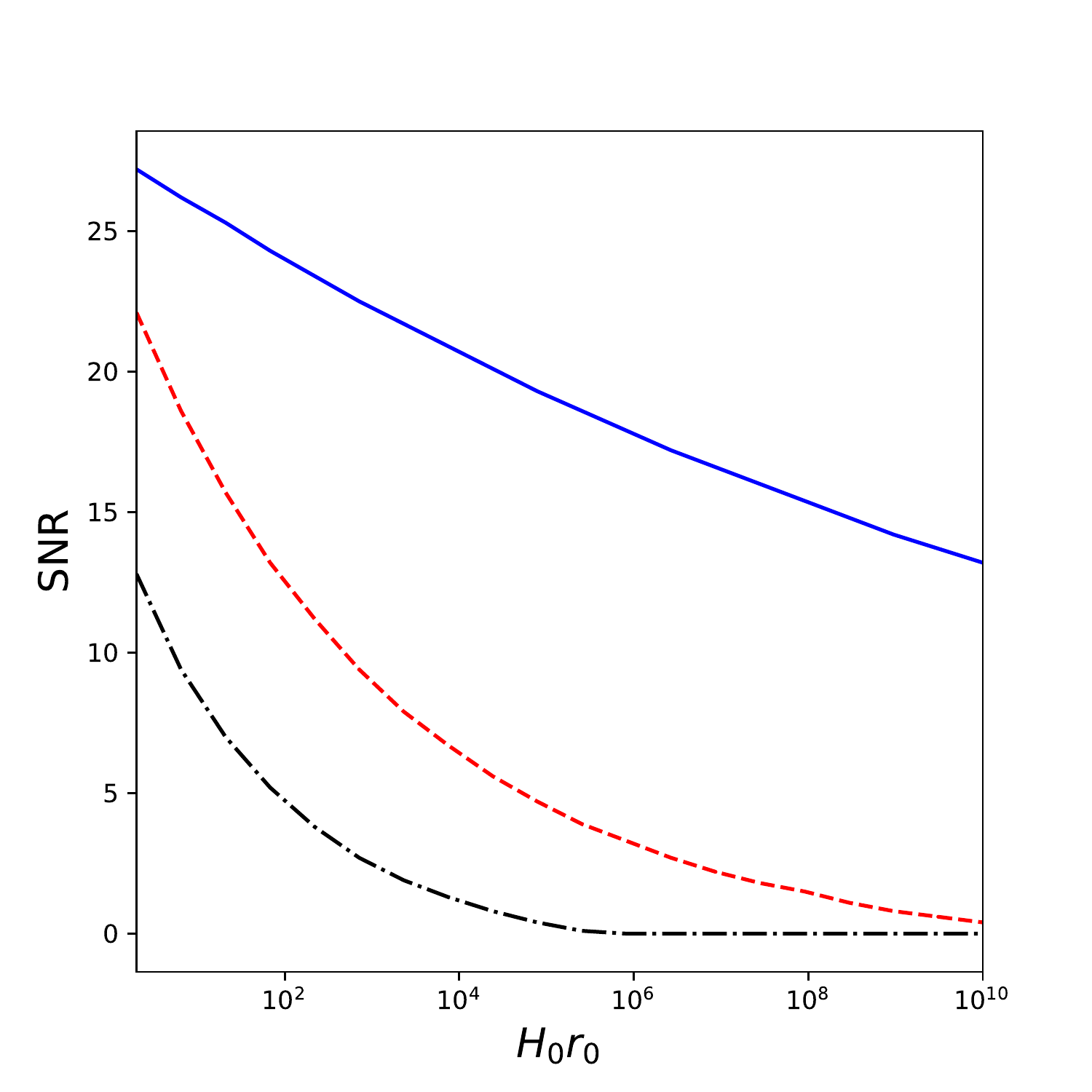}\label{fig:h0r0dep}}
    \caption{The left panel shows the signal to noise ratio for LSST Y10 depending on the lowest multipole measured. A range of values for the spectral indices $n_{\rm DE}$ is explored. We fix the scale of the fluctuations $H_0r_0=10$. In the right panel we show the dependency of the SNR on the scale of the fluctuations for LSST Y10. We consider the same spectral indices. }
    \label{fig:SN}
\end{figure}

The large scales that we want to probe are challenging to extract from the data. On these scales the details of the survey strategy, survey mask and the galactic foreground can make this signal harder to extract. Because of this we have explored how the signal to noise ratio depends on the lowest multipole probed. This is shown in Figure \ref{fig:sn1}. Here we see that for most of the values for $n_{\rm DE}$ a LSST measurement would also be possible for $\ell_{\rm min}\approx20$, but for the case where $n_{\rm DE}=1$ this would be a challenge. In this case it would be worth the effort to constrain these lower multipoles. 

We also studied how the signal to noise ratio depends on the scale $H_0r_0$. This is shown in Figure \ref{fig:h0r0dep}. We observe that as long as the spectral index is not close to one, the signal is measurable even if $H_0 r_0$ becomes relatively large. However, in the case where $n_{\rm DE}=1$ the signal decays rapidly. Measuring this signal when $H_0 r_0>100$ would probably pose a considerable challenge for observationalists. 

Overall, comparing the signal to noise predicted in Figure \ref{fig:nde}, we expect a positive measurement in DES would be very challenging. This is because that even in the case with a very small slope and a relatively small typical length scale of $H_0 r_0=2$, we still find a signal to noise SNR of 1.8. However, due to the very large area of the sky LSST covers, it does not suffer from these effects and could potentially measure even the `harder' case with $H_0 r_0=10$ and $n_{\rm DE}=1$.

\subsection{Redshift dependence}

\begin{figure}
    \centering
    \subfloat[$n_{\rm DE}=0.1$]{\includegraphics[width=0.33\textwidth]{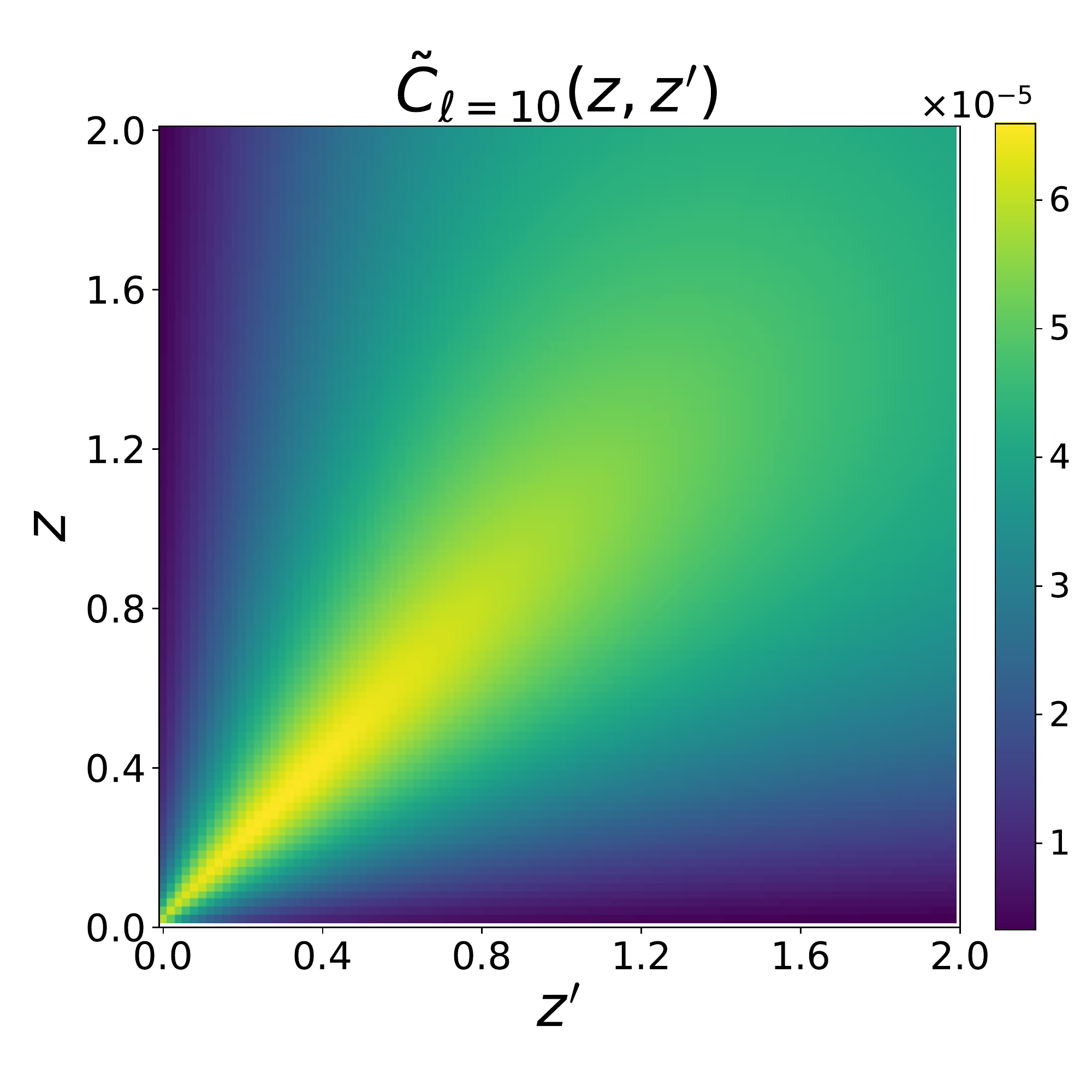}\label{fig:redshift1}}
    \subfloat[$n_{\rm DE}=0.5$]{\includegraphics[width=0.33\textwidth]{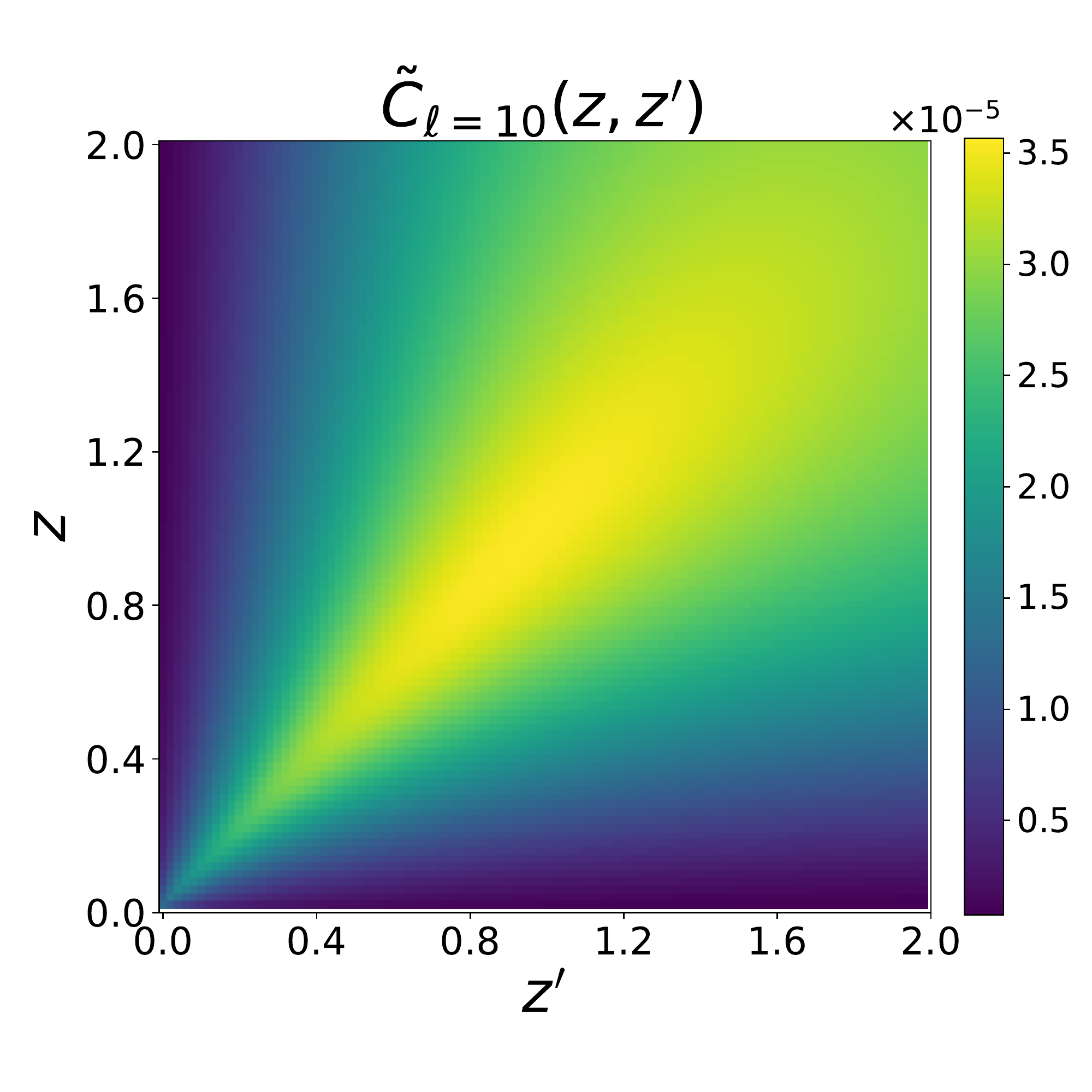}\label{fig:redhsift2}}
    \subfloat[$n_{\rm DE}=1$]{\includegraphics[width=0.33\textwidth]{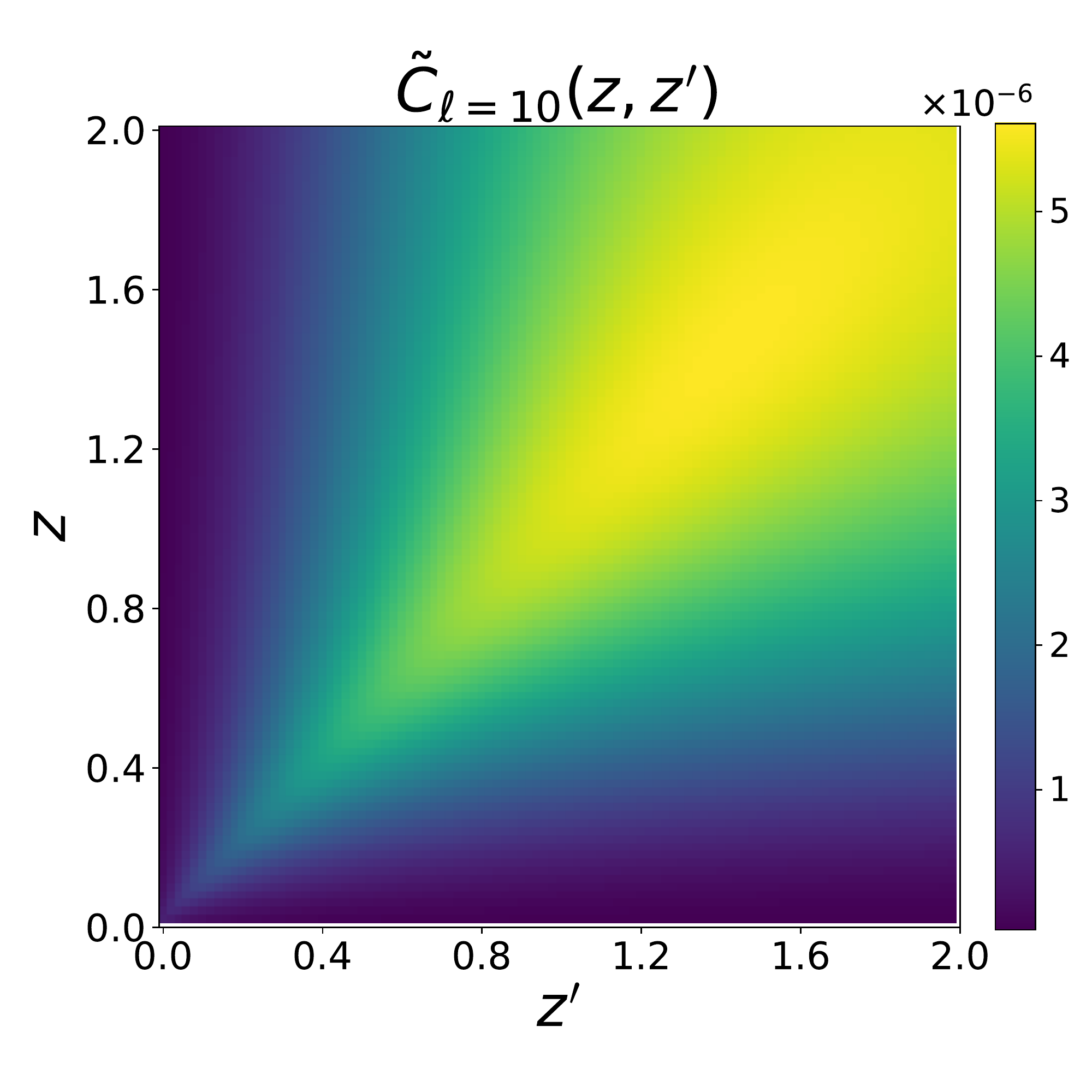}\label{fig:redshift3}}
    \caption{The angular power spectrum (\ref{eq:c(zz)}) as a function redshift $\tilde{C}_{\ell=10}(z,z^\prime)$ shown for several values of the spectral slope $n_{\rm DE}$. These figures are obtained using $H_0 r_0=10$ and $n_{\rm DE}=0.1$ (left), $0.5$ (middle) and $1$ (right). The power spectrum is calculated for the values $z,z^\prime\in(0,2)$ on a $100\times100$ grid. }
    \label{fig:redshift}
\end{figure}

\begin{figure}
    \centering
    \includegraphics[width=0.7\textwidth]{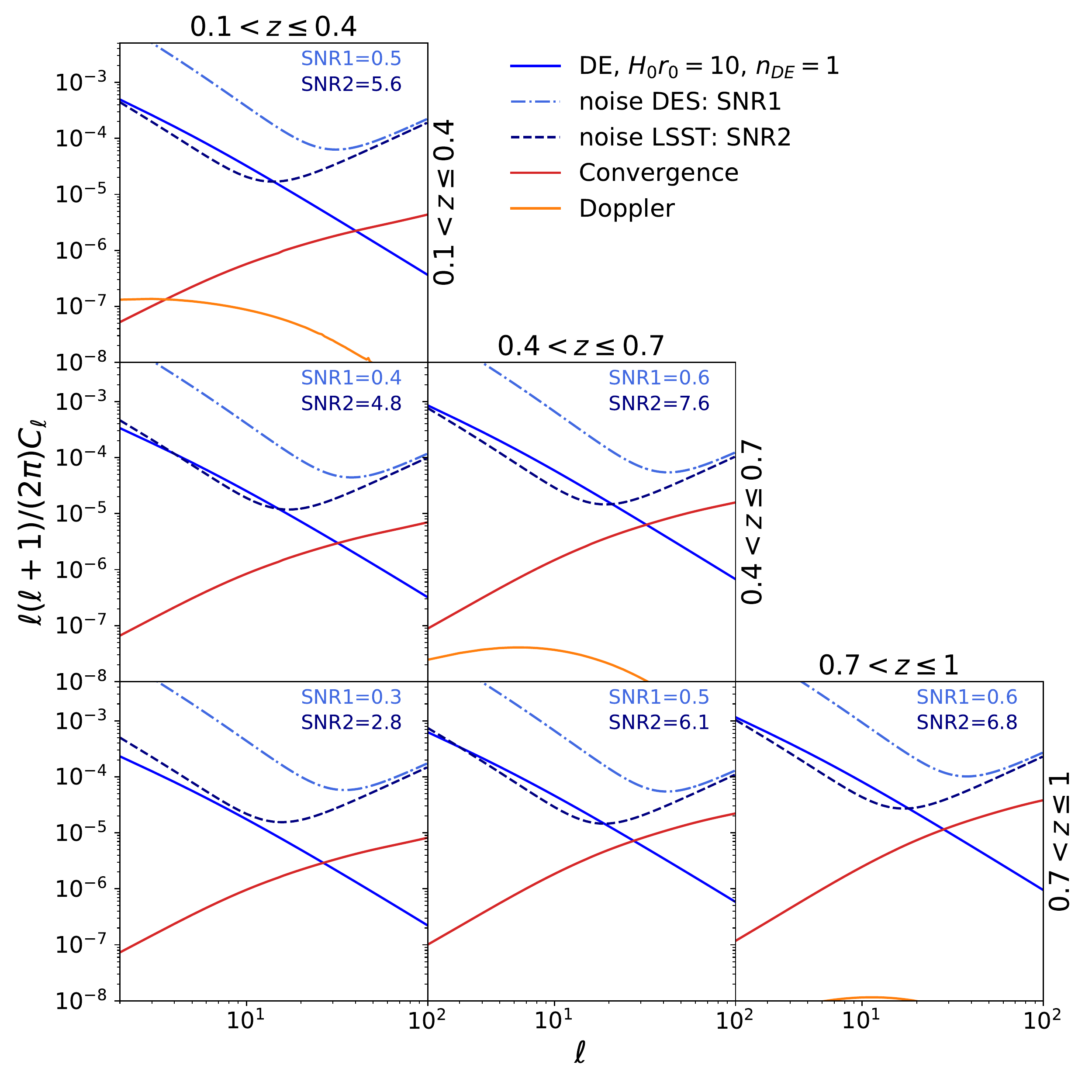}
    \caption{The angular power spectra $C_\ell$ calculated over several redshift bins. We use three bins with width $\Delta z = 0.3$ between $z=0.1$ and $z=1$, using these bins we then consider all auto- and cross-correlations. This then gives the six panels shown above. We also include the lensing effects (convergence - red, Doppler - orange) and the noise expected from both surveys (DES - dot-dashed line, LSST - dashed line). In the top right the expected signal to noise ratio corresponding to this noise is given: SNR1 denotes the signal to noise ratio expected for DES, SNR2 gives the SNR expected for LSST Y10.}
    \label{fig:redshiftbins}
\end{figure}

We perform a study of the redshift dependence of our signal. To this end we show the redshift dependence of $\tilde{C}_{\ell=10}(z,z^\prime)$, which can be calculated from (\ref{eq:c(zz)}). This is shown for several values of the spectral slope $n_{\rm DE}$ in Figure \ref{fig:redshift}. We see that the signal is mostly located at equal redshift, however, it is good to note that the signal does not decay rapidly at unequal redshift. This is also clearly visible in Figure \ref{fig:redshiftbins}, in which we show the signal calculated over several redshift bins, also including cross-correlations between bins. The relatively strong signal at unequal redshift distinguishes itself from the regular cold dark matter power spectrum, which is strongly peaked at equal redshift. This would be beneficial if the value of $H_0 r_0$ is large enough for this effect to be comparable with the Doppler effect. The Doppler effect is  negligible at unequal redshifts, as it requires two objects to be physically close together in order for their velocities to be correlated. This is not true for the convergence spectrum, as this depends on all perturbations along the line of sight and therefore naturally has correlations at unequal redshift. From Figure \ref{fig:redshiftbins} we can also conclude that binning the signal in different redshift bins is not beneficial for the signal to noise, as in every bin the SNR is significantly lower than the SNR obtained for the full distribution (Figure \ref{fig:nde}). We explain this by the fact that the signal has correlations even when considering very different redshifts, therefore considering a broad redshift distribution does not water down the signal. The proper strategy is then reducing the shot noise as much as possible by including the full distribution. However, the unequal redshift signal could be used as a cross-check for the detection. 

From Figure \ref{fig:redshift} we can also see that when the spectral index is smaller, the signal resides more at low redshift. Again, this can be understood from Figure \ref{fig:sr}. Low redshift means probing smaller scales, therefore the part that is most relevant is the small $r$ behaviour in Figure \ref{fig:sr}. A small spectral index means that the function $s(r)$ varies more rapidly when $r$ is small, thereby giving a better signal at low redshift.

\section{Dark Energy of Quantum Origin}\label{sec:quantummodel}
In this section we explore a specific physical model which includes such spatial fluctuations. The scenario that we are considering consists of a quantum field that is spectator during inflation. When quantised, the quantum backreaction of this field sources dark energy at late times. As dark energy has a quantum origin in this case it inherently has spatial fluctuations encoded. 

\subsection{Non minimally coupled scalar field and the stochastic formalism}
This scenario was previously explored by \citep{Glavan2014,Glavan2015,Glavan2016,Glavan2018, Belgacem2021, Belgacem2022}. We follow the approach by \citep{Glavan2018,Belgacem2021,Glavan2016, Belgacem2022} and consider a light scalar field with a non minimal coupling. This is then described by the following action
\begin{equation}\label{eq:lagrangian}
    S[\Phi]=\int d^{4} x \sqrt{-g}\left\{-\frac{1}{2} g^{\mu \nu} \partial_{\mu} \Phi \partial_{\nu} \Phi-\frac{1}{2} m^{2} \Phi^{2}-\frac{1}{2} \xi R \Phi^{2}\right\},
\end{equation}
where $R$ is the Ricci curvature scalar of the metric $g_{\mu \nu}$ and $g=\operatorname{det}\left[g_{\mu\nu}\right]$. It is clear that due to the non minimal coupling this field has an effective mass given by $M^2=m^2 +\xi R$. Due to the Ricci scalar $R$ this is directly dependent on the cosmological background. Previous work by ref. \citep{Glavan2016,Glavan2018} showed that the field-correlators grow during inflation and are subsequently translated into an effective dark energy at late times. This enhancement during inflation is larger when considering the scenario where the non minimal coupling is negative and dominates over the bare mass $m$. The quantum fluctuations of this field are mostly built from super-Hubble modes, so their evolution is frozen after inflation. Therefore these quantum fluctuations can play the role of dark energy.

In this model the fields are quantized in the standard way using canonical quantization. The canonical momentum is ${\Pi}=a^3\dot{\Phi}$, then the fields are promoted to operators obeying the canonical commutation relations.

The effect of the scalar field on the cosmology is determined by the energy momentum tensor, $T_{\mu\nu}(x)= \frac{-2}{\sqrt{-g(x)}}\frac{\delta S_m}{\delta g^{\mu \nu}(x)}$. In operator form this is given by:
\begin{equation}
\hat{T}_{\mu \nu} =\partial_{\mu} \hat{\Phi} \partial_{\nu} \hat{\Phi}-\frac{1}{2} g_{\mu \nu} g^{\alpha \beta} \partial_{\alpha} \hat{\Phi} \partial_{\beta} \hat{\Phi}-\frac{m^{2}}{2} g_{\mu \nu} \hat{\Phi}^{2} 
+\xi\left[G_{\mu \nu}-\nabla_{\mu} \nabla_{\nu}+g_{\mu \nu} \square\right] \hat{\Phi}^{2}.
\end{equation}
$G_{\mu\nu}$ denotes the Einstein tensor and $\square$ is the  box operator or d'Alembertian. With respect to a homogeneous and isotropic state defined on a FLRW background the energy momentum tensor takes the perfect fluid form. From this the energy density can be found. When neglecting spatial gradients this is given by:
\begin{align}
\hat{\rho}_{Q}(t, \vec{x}) &\equiv-\hat{T}_{0}^{0}(t, \vec{x})\\&=\frac{H^{2}}{2}\left\{\left[\left(\frac{m}{H}\right)^{2}+6 \xi\right] \hat{\Phi}^{2}(t, \vec{x})+\frac{6 \xi}{a^{3} H}\{\hat{\Phi}(t, \vec{x}), \hat{\Pi}(t, \vec{x})\}+\frac{1}{a^{6} H^{2}} \hat{\Pi}^{2}(t, \vec{x})\right\}.\label{eq:dens}
\end{align}
The brackets are defined as $\{\hat{A}, \hat{B}\} \equiv \hat{A} \hat{B}+\hat{B} \hat{A}$. Similarly, the pressure is obtained as the $T^i_j$ element of the energy momentum tensor. The problem now reduces to finding the coincident correlators in (\ref{eq:dens}) at late times. This work was done by \citep{Glavan2018} by suitably adopting the stochastic formalism \citep{Starobinsky1986}. In this approach one realises that the dominant contribution to the fields come from super horizon modes ($\mathcal{H}>k$). The field operators are then split into a long and short wavelength part, separated by the comoving scale $\mu \mathcal{H}$. $\mu$ is a UV cutoff, it selects the lowest scale at the beginning of inflation for which the modes are still nearly scale invariant. We take it to be one in our calculations. Hereafter one focuses on the equations of motion for the long wavelengths, where the short wavelength modes enter as a stochastic noise term. 

Using this formalism ref. \citep{Glavan2018} was able to show that the backreaction of this field can indeed result in a viable candidate for dark energy. 

\subsection{Density and Pressure}
In recent work by \citep{Belgacem2022}, it was shown that, assuming a negative non minimal coupling $\xi<0$, in the matter dominated era the density and pressure of these fields can be approximated as follows:
\begin{align}
    \langle \hat{\rho}_{Q}(t, \vec{x}) \rangle&=\frac{\bar{H}^{2}}{2}\left[\left(\frac{m}{\bar{H}}\right)^{2}-6 |\xi|\right] \frac{H_{I}^{2}}{32 \pi^{2}|\xi|} e^{8|\xi| N_{I}+ 4|\xi| N_M},\label{eq:rho} \\
    \langle \hat{p}_Q(t,\vec{x})\rangle &= - \frac{m^2}{2}\frac{H_I^2}{32 \pi |\xi|}e^{8|\xi|N_I+4|\xi|N_M}.
\end{align}
Here, $\bar{H}$ is the Hubble rate during the matter-dominated epoch at time $t$. Due to spatial homogeneity, the expectation value of $\hat{\rho}_Q$ does not depend on the comoving position $\vec{x}$. $H_I$ denotes the constant Hubble rate during inflation, $N_I$ the number of inflationary e-folds, while $N_M$ is the number of e-folds after matter domination. Indeed, the first part of the energy density scales as $\langle\hat{\rho}_Q\rangle=-\langle \hat{p}_Q \rangle$. This is exactly the behaviour of a cosmological constant. Apart from this it also includes a contribution behaving as dark matter with negative energy. Matching the cosmological constant-like term to the observed amount of dark energy today then gives a constraint for the length of inflation, as inflation needs to take long enough to produce the required amount of dark energy \citep{Glavan2018,Belgacem2022}.
\begin{equation}\label{eq:efolds}
    N_{I}=\frac{1}{8|\xi|} \ln \left[24 \pi|\xi|\left(\frac{m_{P}}{H_{I}}\right)^{2}\left(\frac{H_{D E}}{m}\right)^{2}\right]-\frac{N_M}{2},
\end{equation}
where $m_p$ is the Planck mass, $H_{DE}=\sqrt{\Omega_\Lambda} H_0$ and $N_M$ is the number of e-folds starting from matter-radiation equality. $\Omega_\Lambda$ is the fraction of energy density today due to the cosmological constant. For this scenario to unfold, several constraints need to be satisfied:
\begin{equation}\label{eq:contraint}
    |\xi|<\frac{1}{6}\left(\frac{m}{H_{\mathrm{DE}}}\right)^{2}, \quad  \xi<0 \quad \text { and } \quad m / H_{\mathrm{DE}}<1.
\end{equation}
The first constraint is the result from the need that the cosmological constant-like part of $\langle\hat{\rho}_Q\rangle$ has to dominate over the dark matter-like part. The assumptions of light field and negative non-minimal coupling are those which allow a better enhancement of quantum fluctuations \cite{Glavan2018,Belgacem2022} . This can be seen from the effective potential of the scalar field $V(\phi)=\frac{1}{2} M^2 \phi^2$ with $M^2=m^2+6 \xi (2-\epsilon) H^2$. A light field and negative non-minimal coupling are the conditions for which a minimum length of inflation is needed to amplify the quantum fluctuations of the scalar field, which will later manifest as dark energy in matter-dominated epoch and eventually lead the expansion.

The equation for the densities can be rewritten using the $\Omega$ parametrization as
\begin{equation}\label{eq:omeg}
    \Omega_Q(z)=\frac{\langle\hat{\rho}_Q\rangle(z)}{3M_P^2\bar{H}^2(z)}, \qquad \qquad \Omega_c(z)=\frac{\rho_c(z)}{3M_P^2\bar{H}^2(z)}.
\end{equation}
Note that these sum up to one due to the Friedmann equation (\ref{eq:fr}). The redshift dependence of $\langle\hat{\rho}_Q\rangle$ can be derived from (\ref{eq:rho}) and is given by,
\begin{equation}\label{eq:rs}
\frac{\langle\hat{\rho}_Q\rangle(z)}{\langle\hat{\rho}_Q\rangle_0}=\frac{\Omega_\Lambda-\frac{1}{\alpha}\frac{\bar{H}^2(z)}{H_0^2}}{\Omega_\Lambda-\frac{1}{\alpha}},
\end{equation}
the subscript $0$ means the density $\langle\hat{\rho}_Q\rangle$ is evaluated at zero redshift.  We defined $\alpha$ for notational convenience as $\alpha=\frac{1}{6|\xi|}\left(\frac{m}{H_{DE}}\right)^2$. The constraints given by (\ref{eq:contraint}) then require $|\xi|< \frac{1}{6\alpha}$ and $\alpha>1$ for a fixed $\alpha$. Using these expressions, and assuming the classical matter scales as non-relativistic matter we obtain,
\begin{equation}
    1=\Omega_c(z)+\Omega_Q(z)=\Omega_{0,c}(1+z)^3\frac{H_0^2}{\bar{H}^2(z)}+\left[\Omega_\Lambda\frac{H_0^2}{\bar{H}^2(z)}-\frac{1}{\alpha}\right].
\end{equation} 
 This equation can then be solved for the Hubble parameter to give
\begin{equation}\label{eq:fr}
    \frac{\bar{H}^2}{H_0^2} =  \frac{\left(\Omega_M+\frac{1}{\alpha}\right)(1+z)^3+\Omega_\Lambda}{1+\frac{1}{\alpha}},
\end{equation}
where we used that $\Omega_{0,c} = \Omega_M+\frac{1}{\alpha}$, which follows from (\ref{eq:omeg}) and from $\Omega_M=1-\Omega_\Lambda$, where $\Omega_M$ is the matter energy density fraction today. In the limit where $\alpha\rightarrow\infty$, indicating the minimally coupled limit, this indeed reduces to the regular Friedmann equation. Using (\ref{eq:omeg}), (\ref{eq:rs}) and (\ref{eq:fr}) we write down an expression for $\Omega_Q(z)$:
\begin{equation}\label{eq:oqz}
    \Omega_Q(z)=\frac{\Omega_\Lambda-\frac{1}{\alpha}\left(\Omega_M+\frac{1}{\alpha}\right)(1+z)^3}{\Omega_\Lambda+\left(\Omega_M+\frac{1}{\alpha}\right)(1+z)^{3}}.
\end{equation}

\subsection{Spatial correlations}\label{sec:spatcorr}
Since the origin of dark energy in this model is quantum fluctuations of matter fields during inflation, it is expected that this model produces inhomogeneous dark energy of a predictable form. This was later confirmed by ref. \citep{Belgacem2021, Belgacem2022}, which used the same stochastic formalism to calculate the off-coincidence correlators in this theory.

This model produces spatial correlations of the same form as \textit{Case I} using the following identifications \citep{Belgacem2021}, comparing with (\ref{eq:caseI}), we see that the spectral slope can be related to the non minimal coupling by $n_{\rm DE}=16|\xi|$ and the reference scale $r_0$ is connected to the energy scale of inflation by $r_0=\frac{1}{\mu a_I H_I}$:
\begin{equation}
    s(\|\vec{x}-\vec{y}\|) = \begin{cases}3 - 2 \left(\mu a_I H_I r\right)^{16 |\xi|}, \quad \quad \mu a_I H_I r < 1, \\ 1, \quad \quad \quad \quad \quad \quad \quad \quad\quad \  \  \mu a_I H_I r > 1.
    \end{cases}
\end{equation}
Here, $a_I$ and $H_I$ are the scale factor and Hubble parameter at the start of inflation. 

We are mostly interested in the value for the ratio $H_0 r_0$, so we can use (\ref{eq:efolds}) in the following form, where we assumed an instantaneous reheating after inflation \citep{Belgacem2021}: 
\begin{equation}\label{eq:r0h0}
    H_0 r_0=\mu^{-1} e^{N_{I}}\left(\frac{H_{I}}{H_{0}}\right)^{-\frac{1}{2}} \Omega_{R}^{-\frac{1}{4}}.
\end{equation}
Here, $\Omega_R$ is the energy density fraction in radiation today, which we take as $9.1\cdot10^{-5}$. 

Similar to \textit{Case I} in the previous sections, the quantities $\Omega_M$ and $H_0$ are still the unobservable bare quantities. To obtain values for these we match them analogously to the phenomenological model in section \ref{sec:matching}, with the difference being that for $\bar{H}(z)$ and $\Omega_Q(z)$ we now use (\ref{eq:fr}) and (\ref{eq:oqz}). As the redshift dependence is determined by the model parameters, we expand up to the deceleration parameter,
\begin{align}
\frac{1}{H_L}  = & \frac{1}{4 H_0 \alpha^2}\left(3+6 \alpha(\Omega_M-1)+\alpha^2(7+3(\Omega_M-2)\Omega_M))\right), \\
\frac{1}{2 H_L}\left[1-q_L\right]  = & \left(16 H_0 \alpha^2(1+\alpha)\right)^{-1}\left(3+\alpha(42-39\Omega_M)+\alpha^2(3\Omega_M(48-29\Omega_M)-53)+\right.\nonumber \\ & \left. \alpha^3(28-3\Omega_M(27+\Omega_M(15\Omega_M-34))\right)\,,
\end{align}
where we used $\Omega_\Lambda = 1-\Omega_M$. We then solve these equations numerically in the the bare variables $H_0$ and $\Omega_Q$ for several values of $\alpha$. Again, we use the values for $q_L$ and $H_L$ from \citep{Riess_2016}, as explained in \ref{sec:matching}. The results are shown in Table \ref{tab:alphafit}. Values smaller than $\alpha=8$ result in a negative $\Omega_M$, thus we choose $\alpha$ larger than this. Again, we stress that this gives an estimate for the bare parameters, ideally one would fit this model to supernovae data.
\begin{table}
\centering
\begin{tabular}{|l|l|l|l|l|}
\hline
            & $H_0 \left[\mathrm{~km} \mathrm{~s}^{-1} \mathrm{Mpc}^{-1}\right]$ & $\Omega_M$ \\ \hline
$\alpha=10$  & 116.6                                               & 0.03    \\ \hline
$\alpha=25$ & 116.8                                                  & 0.08       \\ \hline
$\alpha=50$ & 116.9                                                 & 0.1        \\ \hline
\end{tabular}
\caption{The values for the bare energy fraction of non-relativistically scaling matter and the bare Hubble constant, calculated for several values of the model constant $\alpha$.}
\label{tab:alphafit}
\end{table}

Using this knowledge we can calculate the $C_\ell$'s. For this model (\ref{eq:finalcl}) takes the following form:
\begin{align}\label{eq:clspec}
    C^{i,j}_\ell =&- {\pi^{\frac{3}{2}}\Omega_{Q,0}^2 H_0^4}\frac{2^{-\ell}\left(-8|\xi|\right)_\ell}{\Gamma(\frac{3}{2}+\ell)}\nonumber\\ &\times\int_0^{\infty} dz \int_0^{\infty} dz^\prime\left\{\frac{W^{i,j}(z,z^\prime)}{\bar{H}(z)^3 \bar{H}(z^\prime)^3}\left[1- (z + z^\prime)\frac{3}{1+\alpha}\frac{1+\alpha\Omega_M}{\alpha(1-\Omega_M)-1}\right] \right.\\ &\left.\times \left(\frac{\chi(z)^2+{\chi(z^\prime)}^2}{r_0^2}\right)^{8|\xi|} \mu_0(z,z^\prime)^{-\ell} {_2}F_1\left(\frac{\ell}{2}-4|\xi|,\frac{1}{2}+\frac{\ell}{2}-4|\xi|;\frac{3}{2}+\ell; \mu_0(z,z^\prime)^{-2}\right)\right\}.\nonumber
\end{align}
\subsection{Prospects for measurability}
\begin{figure}
    \centering
    \subfloat[]{\includegraphics[width=0.5\textwidth]{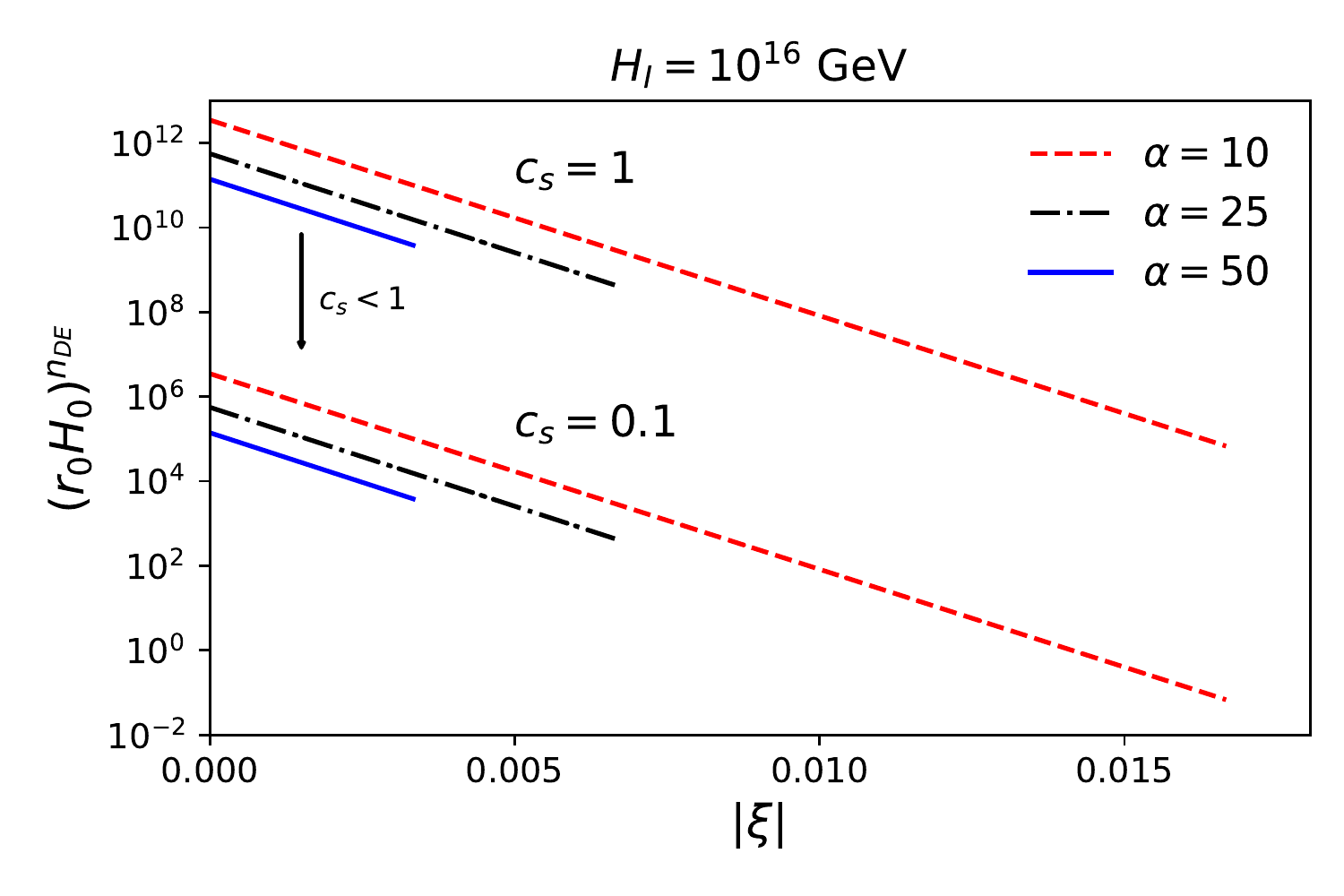}\label{fig:scales}}
    \subfloat[]{\includegraphics[width=0.5\textwidth]{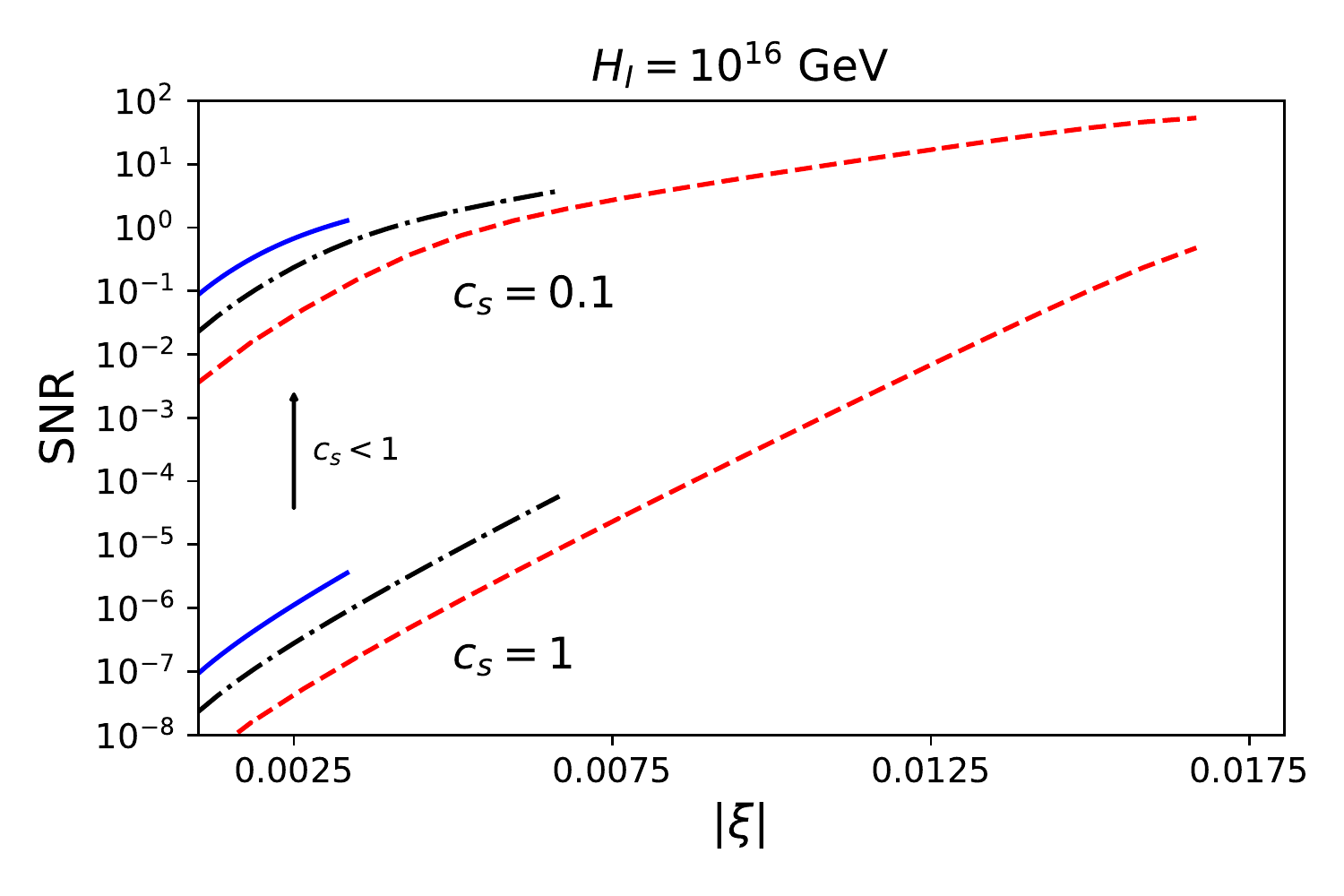}\label{fig:snrqm}}
    \caption{The left panel shows the factor $(H_0 r_0)^{n_{\rm DE}}$ shown as a function of $|\xi|$. The right panel shows the SNR of the model, calculated for the LSST Y10 survey. We keep the scale of inflation fixed at $H_I=10^{16}$ GeV and explore several values of $\alpha$. We also include predictions assuming a speed of sound $c_s=0.1$, which rescales $(H_0 r_0)^{n_{\rm DE}}$ with a factor $c_s^6$. We only show the values that are allowed by the constrained $|\xi|\leq (6\alpha)^{-1}$. }
    \label{fig:qm}
\end{figure}

Before calculating the $C_\ell$'s it is useful to work on the factor $\left(H_0 r_0\right)^{-16 |\xi|}$, as this term can be factored out of the power spectrum, see (\ref{eq:clspec}), and acts as an effective amplitude. Combining (\ref{eq:efolds}) and (\ref{eq:r0h0}) yields the following expression:
\begin{equation}\label{eq:r0h0xi}
    \left(H_0 r_0\right)^{16|\xi|} = \left(\frac{4\pi}{\alpha}\frac{m_p ^2}{H_I^2}\right)^{2}\left(\frac{H_0 \sqrt{\Omega_R}}{H_I \Omega_M }\right)^{8 |\xi|}
\end{equation}
For $H_I$ we adopt the value of $10^{16} \text{ GeV}$, this is around the GUT scale and still below the observational constraint of about $2\times10^{16} \text{ GeV}$ \citep{dodelson2020modern,maggiore2018gravitational}. The signal is only measurable when $(H_0 r_0)^{n_{\rm DE}}$ is not too big, as if this factor becomes too large the signal starts to be drowned in the shot noise. From  (\ref{eq:r0h0xi}) we see that this is mostly determined by the ratio between the Hubble rate today and during inflation, with exponential dependence on $|\xi|$. We also note that $|\xi|$ is limited by $\alpha$ via $|\xi|\leq \frac{1}{6\alpha}$. This turns out to be very restricting. The regime where $(H_0 r_0)^{n_{\rm DE}}$ lies between 2 and $10^4$ mostly lies in the range that is forbidden by this constraint, as can be seen in Figure \ref{fig:scales}. The factor $(H_0 r_0)^{n_{\rm DE}}$ grows exponentially with $|\xi|$, effectively pushing the signal to scales we cannot probe anymore with upcoming surveys such as LSST. This is confirmed by Figure \ref{fig:snrqm}. In this Figure one can observe that in the best case a SNR of 0.46 can be obtained, which means that it would not be measurable by LSST Y10. A smaller $\alpha$ would free up the parameter space to ranges where the signal would be measurable; however, a smaller $\alpha$ leads to a negative $\Omega_M$ in our matching. Alternatively, if the scale of inflation  $H_I$ was higher, it would also be possible to obtain a measurable result. Yet, this range of $H_I$ would already have led to a positive detection of the tensor spectrum, therefore this range is already ruled out by observations. 

From this we conclude the following. To match the results of the $\langle \hat{d}_L \rangle(z)$, the amount of negative matter that this model predicts has to be small, this then leads to a relatively large $\alpha$. To still be able to ensure that the mass of the field $m$ stays small throughout the evolution of the universe, the non minimal coupling $\xi$ also has to be small. Then, to be able to produce enough dark energy to match the amount today, inflation has to take a long time. As our characteristic length scale $H_0 r_0$ grows with the length of inflation this then leads to scales we cannot probe with upcoming experiments.

However, in this analysis we did not take into account the possible effect of a reduced speed of sound $c_s$. A speed of sound $c_s<1$ would reduce $(H_0 r_0)^{16\xi}$. In \cite{Belgacem2022} it was shown that introducing a speed of sound $c_s$ enters as a multiplicative factor of $c_s^6$ to (\ref{eq:r0h0xi}). This enhances the signal significantly, as can be seen in Figure \ref{fig:snrqm}. When $c_s=0.1$, both the cases $\alpha=10$ and $\alpha=25$ are possibly detectable. We note that when the SNR becomes larger, the cosmic variance becomes a relevant noise contribution, resulting in the upper SNR lines exhibiting a bend.

We also point out that the redshift evolution of the model discussed in this section was tested by \citep{Demianski_2019}. They found that this model is slightly favored over $\Lambda$CDM, although not at a statistically significant level.
\section{Conclusion and discussion}
In this paper we explored a simple phenomenological model for dark energy containing spatial correlations and considered its implication for the luminosity distance. We considered two specific instances of this model, related to the way the dark energy density relates to the fundamental quantum fields that build it.  We assume that the energy density of dark energy depends either quadratically (\textit{Case I}) or linearly (\textit{Case II}) on a single underlying field $\hat{\phi}$ with Gaussian statistics. In \textit{Case I} this means the statistics of the density perturbations $\delta \hat{\rho}_Q$ are inherently non-Gaussian.

Our first observation is that in the case that the local variance of the density field is non zero, the expectation value of the luminosity distance becomes dressed by local fluctuations. In the case that the density depends quadratically on the field $\hat{\phi}$, this variance is obtained via Wick's theorem and is naturally large. The result of this is that the locally measured parameters (i.e. $H_0$) might not be equal to the `bare' parameters describing the underlying FLRW universe. The idea that parameters measured do not equal the underlying parameters is not a new idea, as it was previously studied in the context of relativistic universes containing inhomogeneities \citep{Barausse_2005,Bonvin_2006}. But to our knowledge it has to this date not yet been studied from a quantum perspective, where the fluctuations arise due to  quantum effects.

Hereafter we contruct the power spectrum for the fluctuations in the luminosity distance. Using this power spectrum we assess its measurability for DES and LSST, taking into account contaminants from relativistic effects such as the Doppler effect and the convergence effect. From Figure \ref{fig:nde} we find that detecting this signal with DES would be very challenging: the signal is mostly located at the largest scales, making cosmic variance a significant noise contribution. DES only observes a very small patch of the sky, resulting in a large contribution of cosmic variance. On the other hand, LSST will observe the full sky on the southern hemisphere, which greatly reduces the cosmic variance. We then find that depending on the scale of the fluctuations $r_0$ and the spectral index of dark energy $n_{\rm DE}$, defined in (\ref{eq:caseI}), the signal would be measurable. From Figure \ref{fig:h0r0dep} we conclude that when $n_{\rm DE}=1$, the signal is measurable when $H_0r_0 \lesssim 10^2$, while for $n_{\rm DE}=0.1$ the signal would still easily be measurable when $H_0r_0\approx10^{10}$. For the case that $n_{\rm DE}=1$ it would be useful to go through the effort of measuring the lowest multipoles $\ell$, as the signal is mostly located in the lower $\ell$ region (see Figure \ref{fig:sn1}). We also find that the contaminant signal, coming from relativistic effects, does not alter these conclusions. In the regime where these effects become of the same order of our signal, the signal is not measurable as it would be located well below the shot noise contribution (see Figure \ref{fig:nde}). We also find this model has relatively strong correlations at unequal redshift, this could be used as a cross-check for the signal in a tomographic approach. 

With these conclusions in mind we then considered a specific model for dark energy fluctuations \citep{Glavan2014,Glavan2015,Glavan2016,Glavan2018,Belgacem2021,Belgacem2022}. In this model dark energy is linked to a non minimally coupled spectator field during inflation and has subsequently quantum fluctuations imprinted in it. We explore the parameter space of this model. We find that if we want this model to be consistent, the allowed values are such that inflation needs be long (see (\ref{eq:r0h0xi}) and (\ref{eq:efolds})), thereby pushing the signal to scales we cannot probe anymore with upcoming experiments such as LSST, as can be seen in Figure \ref{fig:qm}. Taking into account a speed of sound $c_s<1$ lowers the characteristic length scale $H_0 r_0$ \citep{Belgacem2022} and could thereby make it measurable (see Figure \ref{fig:snrqm}).

To summarize, we found that a fluctuating Hubble rate due to a fluctuating dark energy fluid can have some profound implications for the luminosity distance. As a result of the fluctuating Hubble rate, the luminosity distance itself becomes a fluctuating parameter. Even at the level of the one-point function, the luminosity distance becomes dressed by the fluctuations, resulting in a difference from $\Lambda$CDM. The spatial correlations in the dark energy fluid become visible in the two-point correlator of the luminosity distance. We constructed the angular power spectrum for these fluctuations and studied its detectability.  We find that these fluctuations are mostly visible on very large scales ($\ell\leq50)$ and that for a large part of the models parameter space, this model would be testable by LSST Y10 data. 

We can propose several avenues in continuing this research. In our work we calculated the relativistic effects in the context of perturbed $\Lambda$CDM to obtain an estimate of this effect. Implicitly this assumes that fluctuating dark energy does not alter this effect. A complete treatment would calculate these effects in our model. For this one would need to know how  the gravitational potential is affected by quantum fields containing spatial correlations. This would be an interesting extension of our work. Similarly, it would be interesting to study how the growth of structure would be affected by such models. In the light of the $\sigma_8$ tension, a tension between the CMB \citep{planck2018} and shear measurements \citep{Abbott_2022,Heymans_2021} of $\sigma_8$,  being a major problem in $\Lambda$CDM, we would be interested to see whether it would persist in a model with a fluctuating dark energy candidate. Recently it was shown that such models can alleviate the Hubble tension \citep{Belgacem2021}, which is one of the most pressing issues in current cosmology \citep{Verde_2019,Di_Valentino_2021}. This  shows promise for the $\sigma_8$ tension as well.

In principle, not only dark energy, but also dark matter could have a quantum nature. It would be interesting to see how this would influence causal observables such as the luminosity distance. There has been some progress on the study of dark matter from a fundamental field theoretic perspective \citep{Friedrich_2017,Friedrich_2018,Friedrich_2019}, but as of yet, a study of such effects on the luminosity distance has not been carried out to our knowledge. Another interesting route would be studying how a coupling between dark matter and dark energy would affect our analysis.

We would also be interested in the effect this model has on gravitational waves. In general this would be interesting in the light of upcoming gravitational wave experiments. This would also have another advantage, because the relativistic lensing effects such as lensing both the electromagnetic luminosity distance and the gravitational wave luminosity distance. By considering the difference between the electromagnetic distance and the gravitational wave distance these would thus drop out \citep{Garoffolo2021}. 

The implications of this model on other observables is also a natural route to consider. Observables related to the Hubble parameter would be especially promising, for example, time delays from strongly lensed objects. Other interesting probes would be the cosmic microwave background or lensing studies.

\acknowledgments
This work is part of the Delta ITP consortium, a program of the Netherlands Organisation for Scientific Research (NWO) that is funded by the Dutch Ministry of Education, Culture and Science (OCW) - NWO project number 24.001.027.

\appendix
\section{Non pertubative approach}\label{app:nonp}
In the main text we derived the following expression (\ref{eq: luminositydistance}) for the luminosity distance,
\begin{equation}
    \frac{\langle \hat{d}_L\rangle (z)}{1+z} = \int_0^z d z^\prime \left\langle \frac{1}{\hat{H}(z^\prime,\hat{n})} \right\rangle =  \int_0^z  \frac{dz^\prime}{\bar{H}(z^\prime)}\left\langle \left({{1+\frac{\delta \hat{\rho}_Q(z,\hat{n})}{\rho_{\text{tot}}(z)}}}\right)^{-\frac{1}{2}} \right\rangle.
\end{equation}
We then proceeded with expanding in fluctuations of $\hat{\rho}_Q$. However, this series diverges in both cases, having a zero radius of convergence. To understand until which order we can trust this expansion, and to obtain some insights in the behaviour up to all orders of these fluctuations we derive an exact analytic solution for the one-point functions in both cases by using our assumption that they are fundamentally built out of fields $\hat{\phi}$ obeying Gaussian statistics.

\subsection{Case I}
We assume the density relies on the fields squared, i.e. $\hat{\rho}_Q(z)=B(z)\hat{\phi}^2(z,\hat{n})$, where $B(z)$ is a proportionality constant depending on the specifics of the model. For example, in the quantum origin model discussed previously we have an expression (\ref{eq:dens}) for this constant depending on the model parameters $\xi$ and $m$. 
\begin{figure}
    \centering
    \subfloat[\centering]{\includegraphics[width=0.46\textwidth]{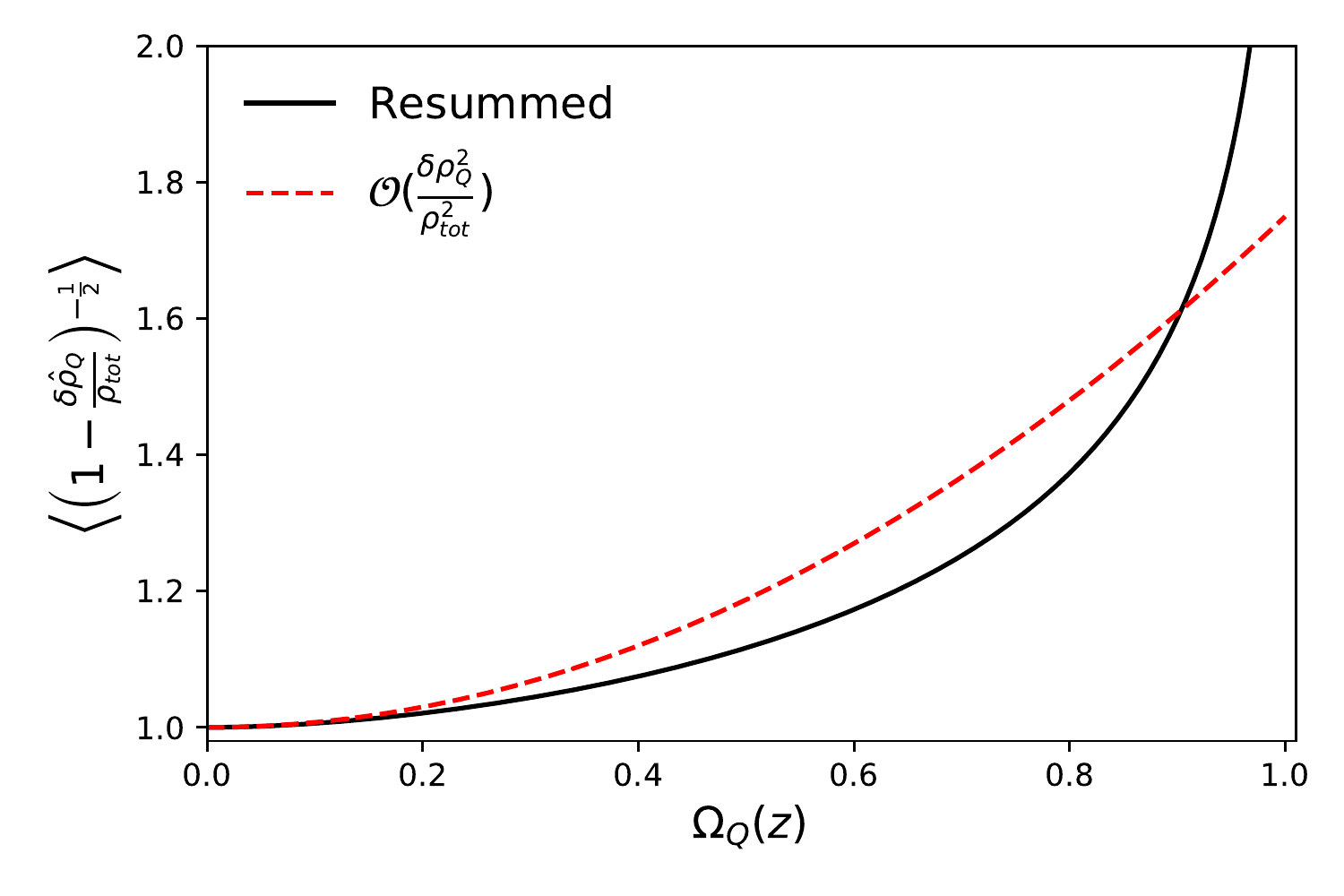}\label{fig:fluc}}
    \subfloat[\centering ]{\includegraphics[width=0.46\textwidth]{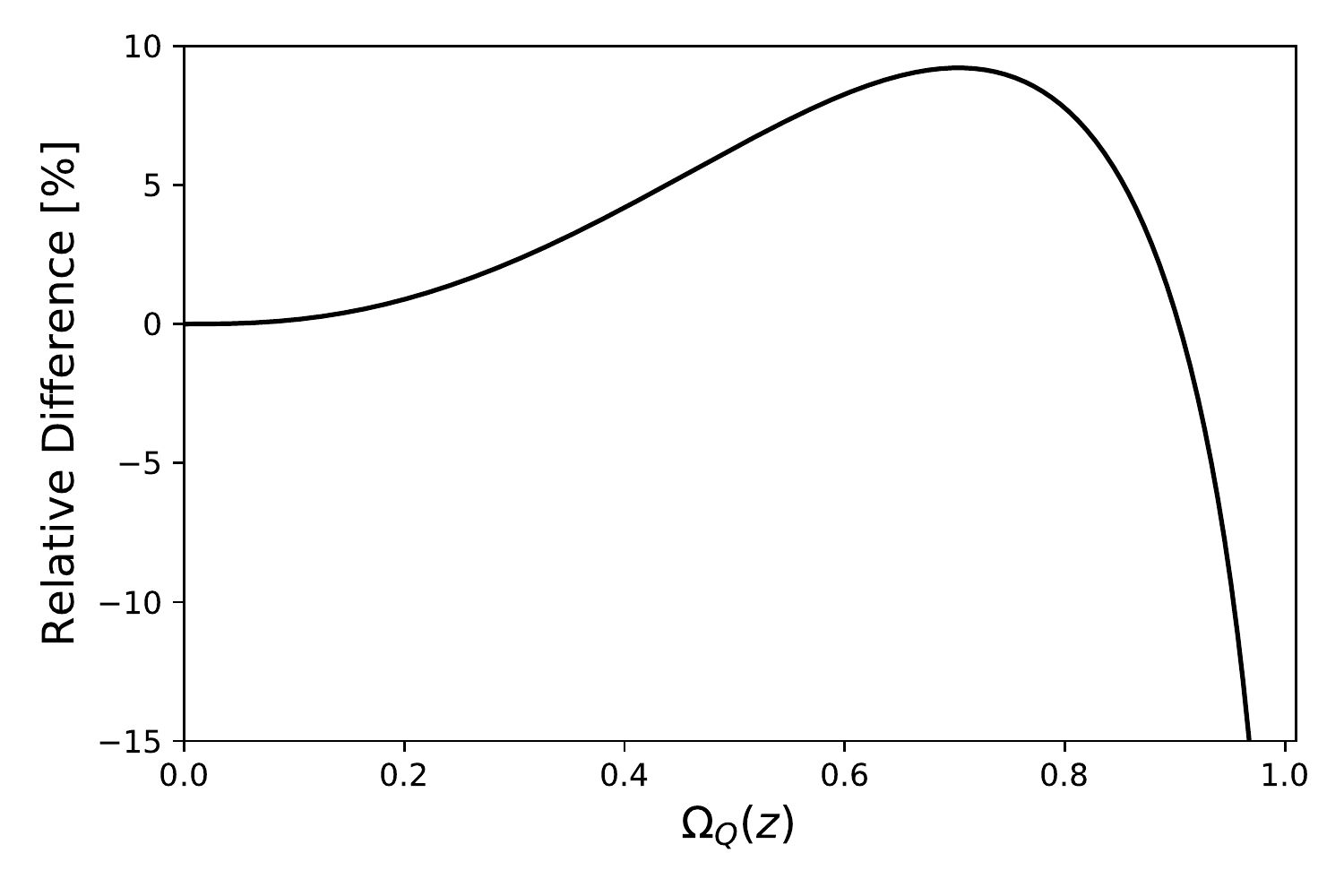}\label{fig:errpr}}
    \caption{The left panel shows both the perturbative ($1+\frac{3}{4}\Omega_Q^2(z)$, red line) and the resummed ((\ref{eq:bes1}),black line) results for the fluctuations in \textit{Case I}. We note that in a universe without perturbations this factor is one. The panel on the right shows the relative difference ($100\times(\operatorname{Perturbative}/\operatorname{Resummed}-1)$) between the perturbative and the resummed results, we observe this to be at most roughly 10 percent.}
    \label{fig:comp}
\end{figure} 
Written in this way, the perturbation written in terms of the fields is given by $\delta \rho_Q(z,\hat{n})=B(z)\left(\hat{\phi}^2(z,\hat{n})-A(z)\right)$, where $A(z)=\langle \hat{\phi}^2 \rangle(z)$. We note that because of this $\delta \hat{\rho}_Q$ obeys non-Gaussian statistics. In the spirit of the stochastic formalism \citep{Starobinsky1986}, we define a classical stochastic variable $\phi$ whose statistical properties are the same as those of the quantum operator $\hat{\phi}$. We know this field obeys Gaussian statistics with vacuum expectation value $\langle\phi\rangle(z)=0$ and variance $\sigma^2(z) = \langle \phi^2 \rangle(z) -\langle\phi\rangle^2(z)=A(z)$. Using this knowledge we can replace the ensemble brackets by an integral over the Gaussian distribution,
\begin{equation}
    \left\langle \left({{1+\frac{\delta \hat{\rho}_Q(z,\hat{n})}{\rho_{\text{tot}}(z)}}}\right)^{-\frac{1}{2}} \right\rangle=\frac{1}{\sqrt{2\pi A(z)}}\int^\infty_{-\infty}  d\phi \frac{e^{-\frac{1}{2}\frac{\phi^2(z)}{A(z)}}}{\sqrt{1+\frac{B(z)}{\rho_\text{tot}(z)}\left(\phi^2(z)-A(z)\right)}}.
\end{equation}
We can make a change of variables $\phi(x) = \sqrt{A(z)} \psi(x)$ and recognize that $\frac{A(z)B(z)}{\rho_\text{tot}(z)}=\frac{\langle\rho_Q\rangle(z)}{\rho_{\text{tot}}(z)}=\Omega_Q(z)$. Then we obtain,
\begin{align}
    \left\langle \left({{1+\frac{\delta \hat{\rho}_Q(z,\hat{n})}{\rho_{\text{tot}}(z)}}}\right)^{-\frac{1}{2}} \right\rangle&=\frac{\sqrt{A(z)}}{\sqrt{2\pi A(z)}}\int^\infty_{-\infty}  d\psi \frac{e^{-\frac{1}{2}\psi^2}}{\sqrt{1+\Omega_Q(z)\left(\psi^2(z)-1\right)}} \\
    &=\frac{1}{\sqrt{2\pi\Omega_Q(z)}}\int_{-\infty}^\infty d\psi \frac{e^{-\frac{1}{2}\psi^2}}{\sqrt{\frac{1}{\Omega_Q(z)}-1+\psi^2}}.
\end{align}
When $\Omega_Q(z) \in [0,1)$ this integral can be solved as a Bessel function yielding,
\begin{equation}\label{eq:bes1}
        \left\langle \left({{1+\frac{\delta \hat{\rho}_Q(z,\hat{n})}{\rho_{\text{tot}}(z)}}}\right)^{-\frac{1}{2}} \right\rangle= \frac{1}{\sqrt{2\pi \Omega_Q(z)}}e^{\frac{1}{4}\left(\frac{1}{\Omega_Q(z)}-1\right)} K_0\left(\frac{1}{4}\left(\frac{1}{\Omega_Q(z)}-1\right)\right).
\end{equation}
$K_0$ denotes the modified Bessel function of the second kind. In Figure \ref{fig:comp} we show both the truncated result at quadratic order in density fluctuations and the resummed result and compare the two. We see that the relative error is at most 10 percent. This justifies using our expansion (\ref{eq:singlecorrelator}), which we used to derive the expectation values for the luminosity distance.

\subsection{Case II}
\begin{figure}
    \centering
    \centering
    \subfloat[\centering]{\includegraphics[width=0.46\textwidth]{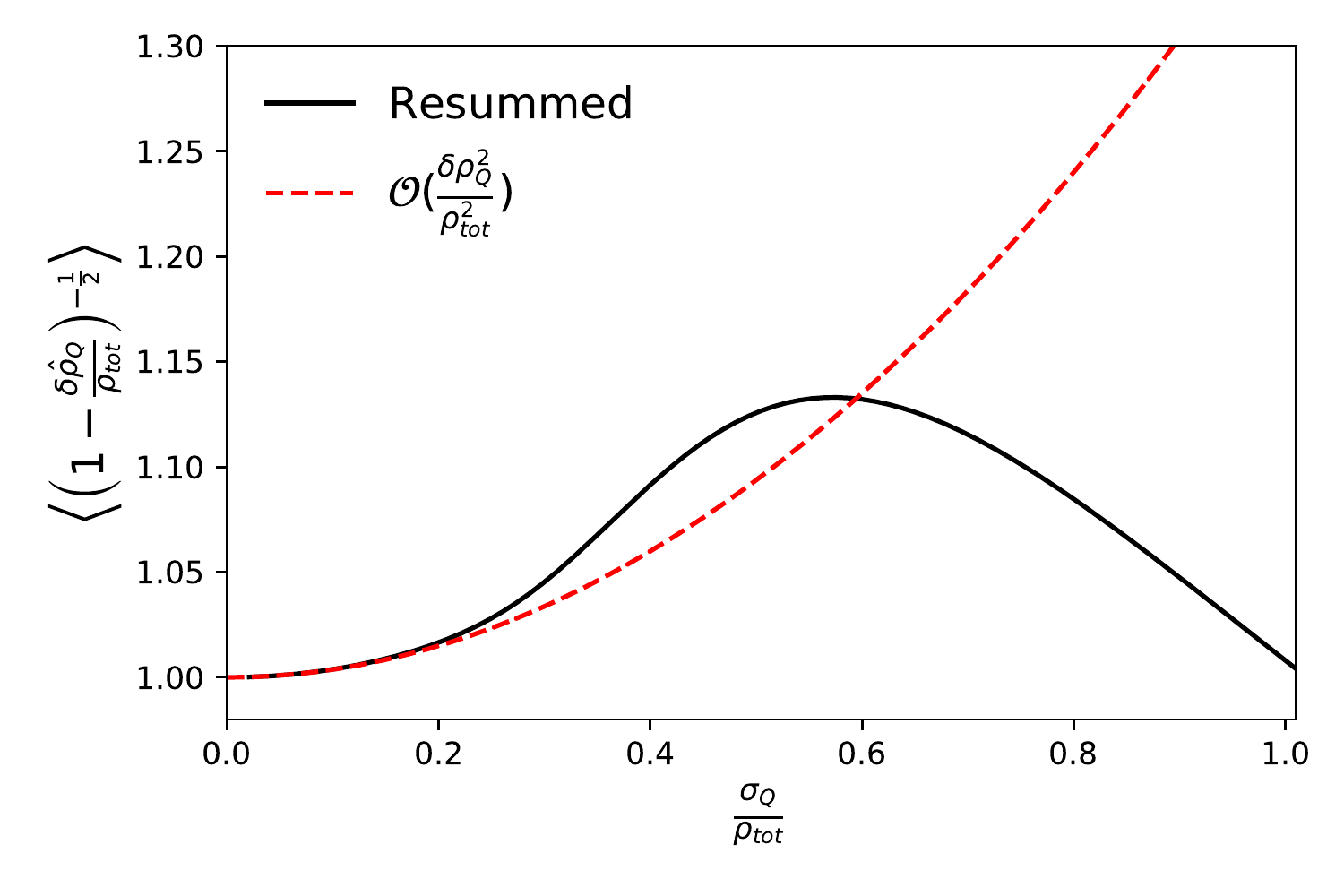}\label{fig:caseIInpcom}}
    \subfloat[\centering ]{\includegraphics[width=0.46\textwidth]{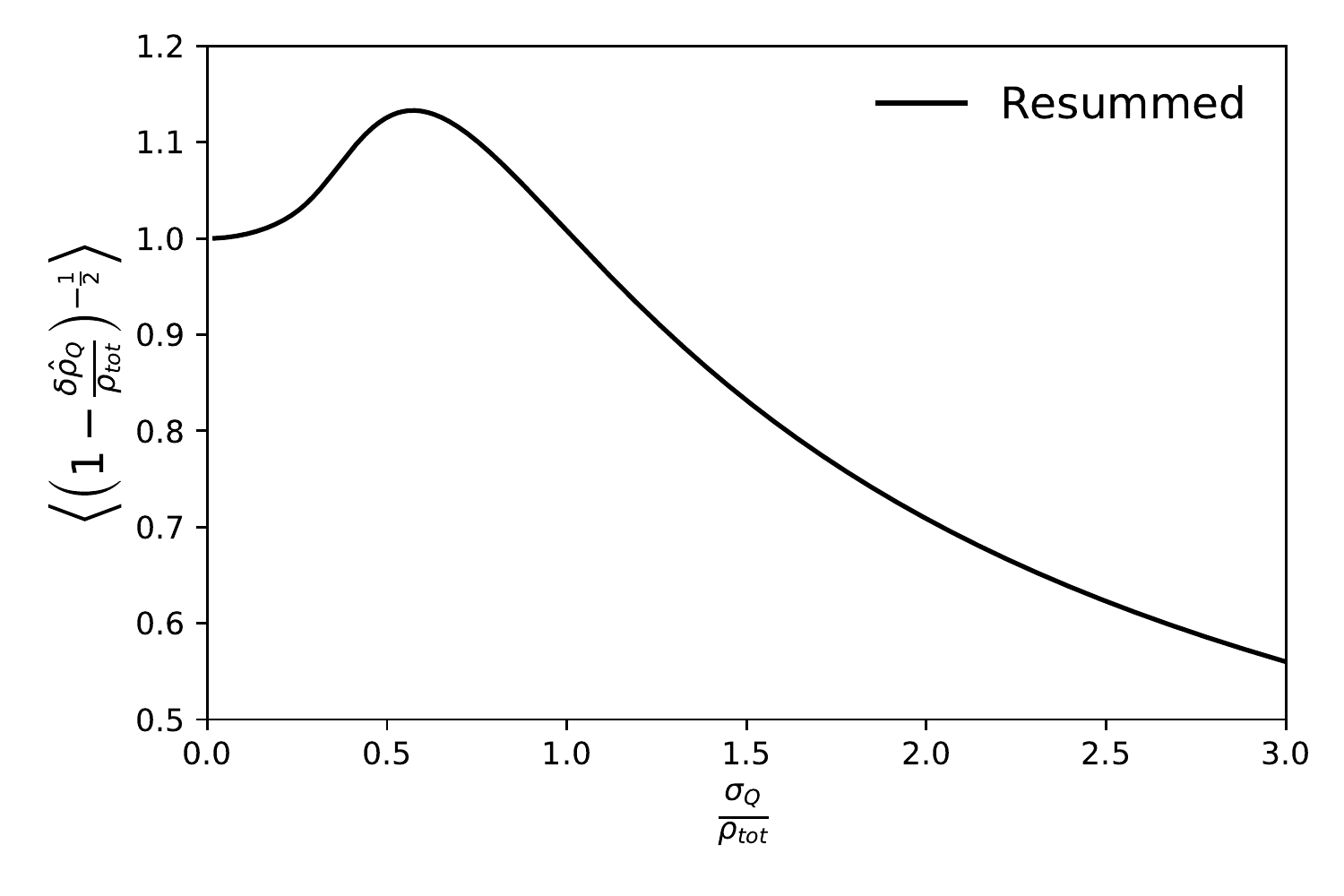}\label{fig:caseIInp}}
    \caption{The term describing the enhancement or decrease due to the fluctuations in the case where the density depends on the fields in a linear fashion (\textit{Case II}). The left panel gives the comparison with our expansion, the black line shows (\ref{eq:bes2}), the red line shows $1+\frac{3}{8}\frac{\sigma_Q^2}{\rho_{tot}^2}$ (see (\ref{eq:singlecorrelator})). The right panel shows (\ref{eq:bes2}) for a large range of $\frac{\sigma_Q}{\rho_{\text{tot}}}$.}
    \label{fig:ok}
\end{figure}
We now extend our treatment to \textit{Case II}, where $\hat{\rho}_Q\propto\hat{\phi}$. Then, when assuming the field $\hat{\phi}$ to be Gaussian, $\delta\hat{\rho}_Q=B(z)\left(\hat{\phi}-\langle\hat{\phi}\rangle\right)$  also obeys Gaussian statistics. We note that this $B(z)$ is a different $B(z)$ than the one in \textit{Case I}. In this case the expectation value is given by,
\begin{equation}
    \left\langle \left({{1+\frac{\delta \hat{\rho}_Q(z,\hat{n})}{\rho_{\text{tot}}(z)}}}\right)^{-\frac{1}{2}} \right\rangle=\frac{1}{\sqrt{2\pi}\sigma}\int_{-\infty}^\infty d\left(\delta\rho\right) \frac{1}{\sqrt{1+\frac{\delta \rho_Q}{\rho_\text{tot}}}}e^{-\frac{1}{2}\frac{\delta \rho_Q^2}{\sigma_Q^2}}.
\end{equation}
Since the RMS Hubble parameter needs to be positive to make physical sense, we change the lower limit. Which is allowed when the Gaussian is sharply peaked, meaning $\frac{\sigma_Q^2}{\rho_{\text{tot}}^2}\ll1$, where $\sigma^2_Q=B(z)^2 \sigma^2$ is the variance of $\hat{\rho}_Q$. This results in:
\begin{equation}
    \left\langle \left({{1+\frac{\delta \hat{\rho}_Q(z,\hat{n})}{\rho_{\text{tot}}(z)}}}\right)^{-\frac{1}{2}} \right\rangle=\frac{1}{\sqrt{2\pi}\sigma_Q}\int_{-\rho_{\text{tot}}}^\infty d\left(\delta\rho_Q\right) \frac{1}{\sqrt{1+\frac{\delta \rho_Q}{\rho_\text{tot}}}}e^{-\frac{1}{2}\frac{\delta \rho_Q^2}{\sigma_Q^2}}.
\end{equation}
 We can first observe that in the case where $\sigma_Q\rightarrow0$ the Gaussian distribution becomes the Dirac delta function $\delta(\delta\rho_Q)$. As a consequence the integral reduces to one, matching the classical result. Now we substitute $x= \frac{\delta\rho_Q}{\rho_\text{tot}}-1$ such that we obtain,
\begin{equation}
     \left\langle \left({{1+\frac{\delta \hat{\rho}_Q(z,\hat{n})}{\rho_{\text{tot}}(z)}}}\right)^{-\frac{1}{2}} \right\rangle =  \frac{1}{\sqrt{{2\pi c}}}  \int_0^\infty dx x^{-\frac{1}{2}}e^{-\frac{ (x-1)^2}{2c}}.
\end{equation}
Here, the constant $c$ is the ratio between the total amount of matter and the fluctuations $c=\left(\frac{\sigma_Q}{\rho_\text{tot}}\right)^2$. This integral can also be solved in terms of Bessel functions, the result is:
\begin{equation}\label{eq:bes2}
    \left\langle \left({{1+\frac{\delta \hat{\rho}_Q(z,\hat{n})}{\rho_{\text{tot}}(z)}}}\right)^{-\frac{1}{2}} \right\rangle = \sqrt{\frac{\pi}{8 }}\frac{\rho_{\text{tot}}}{\sigma_Q}e^{-\frac{\rho^2_{\text{tot}}}{4 \sigma_Q^2}}\left(I_{-\frac{1}{4}}\left(\frac{\rho^2_{\text{tot}}}{4 \sigma^2}\right)+I_{\frac{1}{4}}\left(\frac{\rho^2_{\text{tot}}}{4 \sigma_Q^2}\right)\right).
\end{equation}
Where $I_n$ is the modified Bessel function of the first kind. The behaviour of this function is shown in Figure \ref{fig:caseIInp}. We see that, indeed, it starts at one. This is to be expected: when $\sigma_Q=0$, then the matter does not fluctuate. We also notice that when the dispersion becomes sufficiently large, the fluctuations make the effective Hubble parameter larger, however, in this regime $\frac{\sigma_Q^2}{\rho_{\rm tot}^2}$ is not small so we cannot fully trust this result. In Figure \ref{fig:caseIInpcom} the result is compared with the perturbative result, the agreement is excellent for small $\frac{\sigma_Q^2}{\rho_\text{tot}^2}$. After this ratio surpasses 0.2 the two solutions start to diverge from each other. 

\section{The unequal time correlator}\label{app: unequaltimecorrelator}
In the literature correlators are often calculated at equal time, or equivalently, at equal redshift. In general it is possible to have correlations at different times. It is thus important to calculate the effects of the correlator at different times, a so called `unequal time' correlator. In this Appendix we derive this correlator, which in the main text  is used to obtain (\ref{eq:finalld}).

To obtain such a correlator we expand the equal time correlator around $t=t_0$.
\begin{multline}\label{eq:unequaltimecorrelator}
    \langle \hat{\rho}_Q(t_1, \hat{n}_1) \hat{\rho}_Q(t_2, \hat{n}_2) \rangle = \langle \hat{\rho}_Q(t_0, \hat{n}_1) \hat{\rho}_Q(t_0, \hat{n}_2) \rangle  \\ +  \Delta t_1 \ \langle \dot{\hat{\rho}}_Q(t_0, \hat{n}_1) \hat{\rho}_Q(t_0, \hat{n}_2)\rangle + \Delta t_2\langle \hat{\rho}_Q(t_0, \hat{n}_1) \dot{\hat{\rho}}_Q(t_0, \hat{n}_2)\rangle. 
\end{multline}
With $\Delta t_i = t_i-t_0$ and the dot indicating a partial derivative $\partial_t$. This expansion is good even if $t-t_0$ becomes relatively large: dark energy varies slowly with time and therefore the correlator between $\Dot{\hat{\rho}}_Q$ and $\hat{\rho}_Q$ is small. 

We also write, analogously to (\ref{eq: ansatz}).
\begin{equation}
     \langle \dot{\hat{\rho}}_Q (t,n)\hat{\rho}_Q(t,n^\prime) \rangle = \langle \hat{\rho}_Q (t,n)\hat{\dot{\rho}}_Q(t,n^\prime) \rangle= B(t) \langle \hat{\rho}_Q (t) \rangle \langle \hat{\rho}_Q (t) \rangle s(\|\vec{x}-\vec{y}\|).
\end{equation}

We know that:
\begin{align}
    \partial_t \langle \hat{\rho}_Q (t,n)\hat{\rho}_Q(t,n^\prime) \rangle &=   \langle \dot{\hat{\rho}}_Q (t,n)\hat{\rho}_Q(t,n^\prime) \rangle + \langle \hat{\rho}_Q (t,n)\hat{\dot{\rho}}_Q(t,n^\prime) \rangle \\
    &= 2  B(t) \langle \hat{\rho}_Q (t) \rangle \langle \hat{\rho}_Q (t) \rangle s(\|\vec{x}-\vec{y}\|)\label{eq:b1}.
\end{align}
However, we can also calculate the left hand side by inserting (\ref{eq: ansatz}). We then obtain:
\begin{equation}\label{eq:b2}
    \partial_t \langle \hat{\rho}_Q (t,n)\hat{\rho}_Q(t,n^\prime) \rangle =  \left(2\frac{\langle \dot{\hat{\rho}}_Q(t)\rangle}{\langle \hat{\rho}_Q(t)\rangle}\right) \langle\hat{\rho}_Q\rangle(t)^2 s(\|\vec{x}-\vec{y}\|).
\end{equation}
We can now obtain $B(t)$  by comparing (\ref{eq:b1}) and (\ref{eq:b2}). This gives us:
\begin{equation}
    B(t) =  \frac{\langle \dot{\hat{\rho}}_Q(t)\rangle}{\langle\hat{\rho}_Q(t)\rangle}.
\end{equation}
In total we then get as our result for the correlator in (\ref{eq:unequaltimecorrelator}):
\begin{equation}
    \langle \hat{\rho}_Q(t_1, \hat{n}_1) \hat{\rho}_Q(t_2, \hat{n}_2) \rangle \approx \langle \hat{\rho}_Q(t_0, \hat{n}_1) \hat{\rho}_Q(t_0, \hat{n}_2) \rangle\cdot\left(1+ (\Delta t_1 + \Delta t_2)\frac{\langle\dot{\hat{\rho}}_Q\rangle(t_0)}{\langle\hat{\rho}_Q\rangle(t_0)}\right).\label{eq:uneqcorr}
\end{equation}
\section{A hypergeometric integral}\label{app: hypergeometricintegral}
We are interested in solving the following integral, which appears in our expression for the angular power spectrum (\ref{eq:clint}),
\begin{equation}
    I=\int_{-1}^{1} d\mu \left( \frac{r(z_1,z_2,\mu)^{n_{\rm DE}}}{r_0^{n_{\rm DE}}}\right)\mathcal{P}_\ell(\mu),
\end{equation}
with $r$ being the comoving distance between two points. We can rewrite $r(z_1,z_2,\mu)^{n_{\rm DE}}=\left(\chi_1^2+\chi^2_2-2\chi_1\chi_2\mu\right)^{\frac{{n_{\rm DE}}}{2}}$, where $\chi_i=\chi(z_i)$, the comoving distance to the point $z_i$.  We can then rewrite the integral as
\begin{equation}
    I = \left(2\chi_1\chi_2\right)^{\frac{{n_{\rm DE}}}{2}}\frac{\mu_0^\frac{{n_{\rm DE}}}{2}}{r_0^{n_{\rm DE}}}\int_{-1}^1 d\mu\left(1-\frac{\mu}{\mu_0}\right)^\frac{{n_{\rm DE}}}{2}\mathcal{P}_\ell(\mu) =  \frac{\left(2\chi_1\chi_2\right)^{\frac{{n_{\rm DE}}}{2}}}{r_0^{n_{\rm DE}}}\tilde{I}.
\end{equation}
Here, we defined $\tilde{I}$ as the integral over $\mu$ and $\mu_0=\frac{\chi_1^2+\chi_2^2}{2\chi_1\chi_2}$. As the ratio $\frac{\mu}{\mu_0}$ is smaller than one we can write it as a series using Newtons binomium, then commute the sum and the integral,
\begin{equation}
   \tilde{I} =  \mu_0^{\frac{{n_{\rm DE}}}{2}} \int_{-1}^1 d\mu (1-\frac{\mu}{\mu_0})^{\frac{{n_{\rm DE}}}{2}}\mathcal{P}_\ell(\mu) = \sum_{n=0}^\infty  \binom{\frac{{n_{\rm DE}}}{2}}{n}\frac{1}{(-\mu_0)^n} \int_{-1}^1 d\mu \mu^n \mathcal{P}_\ell(\mu)\label{eq:gamsum},
\end{equation}
 We have now replaced our integral with a summation and a more tractable integral. A very similar integral is given in ref.  \cite{gradshteyn2014}, namely:
\begin{equation}
     \int_{0}^1 d\mu x^n \mathcal{P}_\ell(\mu) = \frac{\sqrt{\pi}}{2^{1+n}}\frac{\Gamma(1+n)}{\Gamma(1+\frac{n-\ell}{2})\Gamma(\frac{3}{2}+\frac{n+\ell}{2})}.
\end{equation}
Here, $\Gamma(z)$ is Eulers Gamma function. To calculate the integral on the other half of our  domain we make use of the fact that $\mathcal{P}_\ell(-x) = (-1)^\ell \mathcal{P}_\ell(x)$. For the complete domain the integral is then given by
\begin{equation}
    \int_{-1}^1 d\mu x^n \mathcal{P}_\ell(\mu)=\left(1+(-1)^{\ell+n}\right)\frac{\sqrt{\pi}}{2^{1+n}}\frac{\Gamma(1+n)}{\Gamma(1+\frac{n-\ell}{2})\Gamma(\frac{3}{2}+\frac{n+\ell}{2})}.
\end{equation}
We can insert this in (\ref{eq:gamsum}), yielding
\begin{equation}\label{eq:gamsum2}
    \tilde{I} = \frac{\mu_0^{\frac{{n_{\rm DE}}}{2}}\sqrt{\pi}}{2} \sum_{n=0}^\infty  \binom{\frac{{n_{\rm DE}}}{2}}{n}\left(1+(-1)^{\ell+n}\right)\left(-\frac{1}{ 2\mu_0}\right)^n\frac{\Gamma(1+n)}{\Gamma(1+\frac{n-\ell}{2})\Gamma(\frac{3}{2}+\frac{n+\ell}{2})}.
\end{equation}
Now, we rewrite the binomial coefficient to obtain:
\begin{equation}
    \tilde{I} = \frac{\mu_0^{\frac{{n_{\rm DE}}}{2}}\sqrt{\pi}}{2 \Gamma(-\frac{{n_{\rm DE}}}{2})} \sum_{n=0}^\infty \left(1+(-1)^{\ell+n}\right)\left(\frac{1}{ 2\mu_0}\right)^n\frac{\Gamma(n-\frac{{n_{\rm DE}}}{2})}{\Gamma(1+\frac{n-\ell}{2})\Gamma(\frac{3}{2}+\frac{n+\ell}{2})}.
\end{equation}
Our goal is to rewrite this in terms of Pochammer symbols, $(a)_n = \frac{\Gamma(a+n)}{\Gamma(a)}$, which do not include factors $\frac{n}{2}$. However, we can first consider the even part of this integral:
\begin{equation}
    \tilde{I}_{\text{even}} = \frac{\mu_0^{\frac{{n_{\rm DE}}}{2}}\sqrt{\pi}}{ \Gamma(-\frac{{n_{\rm DE}}}{2})} \sum_{n=0}^\infty \left(\frac{1}{ 2\mu_0}\right)^{2n}\frac{\Gamma(2n-\frac{{n_{\rm DE}}}{2})}{\Gamma(1+n-\frac{\ell}{2})\Gamma(\frac{3}{2}+n+\frac{\ell}{2})}.
\end{equation}
We have now gotten rid of the factors $n/2$, but we have obtained factors $2 n$. To remove these we use Gauss' multiplication theorem to rewrite these in terms of Gamma function with factors $n$. This yields:
\begin{equation}
    \tilde{I}_{\text{even}} = \frac{\mu_0^{\frac{{n_{\rm DE}}}{2}}2^{-\frac{{n_{\rm DE}}}{2}-1}}{ \Gamma(-\frac{{n_{\rm DE}}}{2})} \sum_{n=0}^\infty \left(\frac{1}{ \mu_0^2}\right)^{n}\frac{\Gamma(n-\frac{{n_{\rm DE}}}{4})\Gamma(n-\frac{{n_{\rm DE}}}{4}+\frac{1}{2})}{\Gamma(1+n-\frac{\ell}{2})\Gamma(\frac{3}{2}+n+\frac{\ell}{2})}
\end{equation}
Now we may note that $\Gamma(1+n-\frac{\ell}{2})$ hits a pole if $\frac{\ell}{2}\geq 1+n$, which means that all terms up to and including $n=\frac{\ell}{2}-1$ do not contribute. This allows us to relabel the starting index of the summation to $n=\frac{\ell}{2}$, after this we can relabel everything to make the summation start at 0 again. This yields
\begin{align}\label{eq:sumg}
    \tilde{I}_{\text{even}}&= \frac{\mu_0^{\frac{{n_{\rm DE}}}{2}-\ell}2^{-\frac{{n_{\rm DE}}}{2}-1}}{ \Gamma(-\frac{{n_{\rm DE}}}{2})} \sum_{n=0}^\infty \frac{\left(\frac{1}{ \mu_0^2}\right)^{n}}{n!}\frac{\Gamma(n+\frac{\ell}{2}-\frac{{n_{\rm DE}}}{4})\Gamma(n+\frac{\ell}{2}-\frac{{n_{\rm DE}}}{4}+\frac{1}{2})}{\Gamma(\frac{3}{2}+n+\ell)}.
\end{align}
We can now recognise Gauss' hypergeometric function in (\ref{eq:sumg}). Therefore we can write the integral as
\begin{multline}
    \tilde{I}=\frac{\mu_0^{\frac{{n_{\rm DE}}}{2}-\ell}2^{-\frac{{n_{\rm DE}}}{2}-1}}{ \Gamma(-\frac{{n_{\rm DE}}}{2})}\frac{\Gamma(\frac{\ell}{2}-\frac{{n_{\rm DE}}}{4})\Gamma(\frac{1}{2}+\frac{\ell}{2}-\frac{{n_{\rm DE}}}{4})}{\Gamma(\frac{3}{2}+\ell)}\\\times {_2}F_1\left(\frac{\ell}{2}-\frac{{n_{\rm DE}}}{4},\frac{1}{2}+\frac{\ell}{2}-\frac{{n_{\rm DE}}}{4};\frac{3}{2}+\ell;\frac{1}{\mu_0^2}\right),
\end{multline}
where we dropped the even label, as going through the same steps in the odd case yields the same answer. For the total integral we have:
\begin{multline}
    I = \frac{\left(2\chi_1\chi_2\right)^{\frac{{n_{\rm DE}}}{2}}}{r_0^{n_{\rm DE}}}\sqrt{\pi}\mu_0^{\frac{{n_{\rm DE}}}{2}-\ell}2^{-\ell}\frac{\Gamma\left(-\frac{{n_{\rm DE}}}{2}+\ell\right)}{\Gamma(-\frac{{n_{\rm DE}}}{2})\Gamma(\frac{3}{2}+\ell)}\\ \times {_2}F_1\left(\frac{\ell}{2}-\frac{{n_{\rm DE}}}{4},\frac{1}{2}+\frac{\ell}{2}-\frac{{n_{\rm DE}}}{4};\frac{3}{2}+\ell;\frac{1}{\mu_0^2}\right),
\end{multline}
where we have used Gauss' multiplication theorem to simplify the prefactor slightly.

\section{Angular and ensemble averaging}\label{app:ergodic}
We derived an expression for the residuals of the luminosity distance, however, we obtained this expression in terms of ensemble averages. In practice observations are done over the past light cone, invoking an angular average. The relationship between these two averages is in principle complicated and depends on the specific observable \citep{Bonvin_2015,Fleury_2017,Yoo_2019a,Yoo_2019b}. Differences between these two can have several origins; a state average is for example defined over a space like hypersurface of constant time, while the angular average is an average over the past light cone. However, we focus on another aspect, namely the fact that due to cosmic variance the average we observe is different from the true background average. The true background average is in principle not measurable, due to the fact that we can only access our past light cone. We, therefore, have no way of knowing if our measured average is the true one, as for this we would have to do different measurements at different positions. This problem was discussed in the context of the Cosmic Microwave Background by  \citep{Yoo_2019a}. We extend this treatment and apply it to our own case.

The problem becomes clear when considering the observed luminosity distance:
\begin{equation}
    \bar{d}_L = \int \frac{d^2 \hat{n}}{4\pi} \hat{d}_L(z,\hat{n}),
\end{equation}
this is still a stochastic quantity: a measurement on a different location could give a different value, as our measurement could be influenced by fluctuations. Formally we can write this statement as $\bar{d}_L=\langle \hat{d}_L \rangle+\left(\bar{d}_L - \langle \hat{d}_L \rangle\right)$, where the term in brackets is not necessarily zero. We can now define the cosmic variance $\delta_{cv} = \frac{\bar{d}_L}{\langle \hat{d}_L \rangle}-1$. If $\delta_{cv}=0$, then the two averages are the same. In terms of this cosmic variance we have $\bar{d}_L=\langle \hat{d}_L \rangle (\delta_{cv} +1)$. 

As we are interested in the observational accessible average, we want to make predictions for the fluctuations with respect to the angular mean $\tilde{\Delta}$, while theoretically we have predictions for the fluctuations with respect to the ensemble average $\Delta$. These are defined as:
\begin{equation}
\tilde{\Delta}(z,\hat{n}) = \frac{\hat{d}_L}{\bar{d}_L}-1, \quad \quad \quad \quad {\Delta}(z,\hat{n}) = \frac{\hat{d}_L}{\langle{d}_L \rangle}-1.
\end{equation}
Fluctuations become visible in the two-point functions, for these two quantities these are given by:
\begin{align}
    \langle \tilde{\Delta}(z,\hat{n})\tilde{\Delta}(z^\prime,\hat{m})\rangle &= \frac{\langle \hat{d}_L(z,\hat{n})\hat{d}_L(z^\prime,\hat{m})\rangle}{\bar{d}_L(z) \bar{d}_L(z^\prime)}-1, \\     \langle {\Delta}(z,\hat{n}){\Delta}(z^\prime,\hat{m})\rangle &= \frac{\langle \hat{d}_L(z,\hat{n})\hat{d}_L(z^\prime,\hat{m})\rangle}{\langle{d}_L \rangle(z) \langle{d}_L\rangle(z^\prime)}-1.
\end{align}
The task at hand is then expressing the first quantity   in terms of the theoretically available second quantity. To this end we can express $\Delta$ in terms of $\tilde{\Delta}$ and the cosmic variance $\delta_{cv}$,
\begin{equation}
    \Delta(z,\hat{n}) = \tilde{\Delta}(z,\hat{n})+\delta_{cv}(z)+\tilde{\Delta}(z,\hat{m}) \delta_{cv}(z)
\end{equation}
Using this we can express the ensemble correlator in terms of correlation functions including $\tilde{\Delta}$ and $\delta_{cv}$:
\begin{align}
    \langle \Delta(z,\hat{n}) \Delta(z^\prime,\hat{m}) \rangle =& \langle \tilde{\Delta}(z,\hat{n})\tilde{\Delta}(z^\prime,\hat{m}) \rangle + \langle \tilde{\Delta}(z,\hat{n}) \delta_{cv}(z^\prime) \rangle + \langle \tilde{\Delta}(z^\prime,\hat{m}) \delta_{cv}(z) \rangle\nonumber \\& + \langle \delta_{cv}(z) \delta_{cv}(z^\prime) \rangle +\langle \tilde{\Delta}(z,\hat{n}) \delta_{cv}(z^\prime) \tilde{\Delta}(z^\prime,\hat{m})\rangle  \\ &+ \langle \tilde{\Delta}(z^\prime,\hat{m}) \delta_{cv}(z)\tilde{\Delta}(z,\hat{n})\rangle\nonumber + \langle \delta_{cv}(z) \tilde{\Delta}(z,\hat{n})\delta_{cv}(z^\prime) \rangle\\ &+ \langle \delta_{cv}(z^\prime) \tilde{\Delta}(z^\prime,\hat{m})\delta_{cv}(z) \rangle  + \langle\tilde{\Delta}(z,\hat{n})\delta_{cv}(z) \tilde{\Delta}(z^\prime,\hat{m}) \delta_{cv}(z^\prime)\rangle.\nonumber
\end{align}
 We now assume Wick's theorem, Wick's theorem is valid for Gaussian distributed functions. Which we take to be approximately true. Because $\langle \tilde{\Delta}\rangle=\langle\delta_{cv}\rangle=0$ the three-point functions vanish. The four-point function can be expressed in three products of two-point functions. This yields:
\begin{align}
    \langle \Delta(z,\hat{n}) \Delta(z^\prime,\hat{m}) \rangle \approx& \langle \tilde{\Delta}(z,\hat{n})\tilde{\Delta}(z^\prime,\hat{m}) \rangle + \langle \tilde{\Delta}(z,\hat{n}) \delta_{cv}(z^\prime) \rangle + \langle \tilde{\Delta}(z^\prime,\hat{m}) \delta_{cv}(z) \rangle\nonumber  \\& + \langle \delta_{cv}(z) \delta_{cv}(z^\prime) \rangle + \langle \delta_{cv}(z^\prime) \tilde{\Delta}(z^\prime ,\hat{m})\rangle \langle \delta_{cv}(z) \tilde{\Delta}(z,\hat{n})\rangle\label{eq:expr} \\ & + \langle \delta_{cv}(z^\prime) \tilde{\Delta} (z,\hat{n})\rangle \langle \delta_{cv}(z) \tilde{\Delta}(z^\prime,\hat{m})\rangle  + \langle\delta_{cv}(z) \delta_{cv}(z^\prime) \rangle \langle \tilde{\Delta}(z,\hat{n})\tilde{\Delta}(z^\prime,\hat{m}) \rangle.\nonumber
\end{align}
The remaining task is now finding the suitable expressions for the correlators involving the cosmic variance, $\langle \delta_{cv}(z)\delta_{cv}(z^\prime) \rangle$ and $\langle \tilde{\Delta}(z,\hat{n})\delta_{cv}(z^\prime) \rangle$. We start with the auto correlator, we expand the definition to obtain:
\begin{equation}
    \langle \delta_{cv}(z) \delta_{cv}(z^\prime) \rangle = \left\langle\left(\frac{\bar{d}_L(z)}{\langle \hat{d}_L \rangle(z)}-1 \right)\left(\frac{\bar{d}_L(z^\prime)}{\langle \hat{d}_L^\prime \rangle(z^\prime)}-1 \right)\right\rangle = \frac{\langle \bar{d}_L(z) \bar{d}_L(z^\prime) \rangle}{\langle \hat{d}_L \rangle (z) \langle \hat{d}_L \rangle (z^\prime)}-1.
\end{equation}
We can now factor the angular averaging integrals out of the ensemble brackets. This gives us:
\begin{align}
    \langle \delta_{cv}(z) \delta_{cv}(z^\prime) \rangle =& \int \frac{d^2\hat{n}}{4\pi} \int \frac{d^2 \hat{m}}{4\pi}\left[ \frac{\langle \hat{d}_L(z,\hat{n}) \hat{d}_L(z^\prime,\hat{m}) \rangle}{\langle \hat{d}_L \rangle (z) \langle \hat{d}_L \rangle (z^\prime)}-1\right]\label{eq:cvmono}\\ =& \frac{1}{2}\int_{-1}^1 d\mu\left[ \frac{\langle \hat{d}_L(z) \hat{d}_L(z^\prime) \rangle(\mu)}{\langle \hat{d}_L \rangle (z) \langle \hat{d}_L \rangle (z^\prime)}-1\right] \\ =& \frac{C_0(z,z^\prime)}{4\pi}\label{eq:cvmono1},
\end{align}
where we used the fact that the correlator only depends on the relative angle. This angle is contained in $\mu=\hat{n}\cdot\hat{m}=\cos(\theta)$ with $\theta$ being the relative angle. In the last step we recognise the angular power spectrum (\ref{eq:cls}). This is an expression we can calculate and we thus have a prediction for the auto correlator of the cosmic variance. 

Inserting this in (\ref{eq:expr}) we obtain:
\begin{align}
    \langle \Delta(z,\hat{n}) \Delta(z^\prime,\hat{m}) \rangle \approx & \langle \tilde{\Delta}(z,\hat{n})\tilde{\Delta}(z^\prime,\hat{m}) \rangle\left(1+\frac{C_0(z,z^\prime)}{4\pi}\right) + \langle \tilde{\Delta}(z,\hat{n}) \delta_{cv}(z^\prime) \rangle \nonumber \\&+ \langle \tilde{\Delta}(z^\prime,\hat{m}) \delta_{cv}(z) \rangle +\frac{C_0(z,z^\prime)}{4\pi}+ \langle \delta_{cv}(z^\prime) \tilde{\Delta}(z^\prime ,\hat{m})\rangle \langle \delta_{cv}(z) \tilde{\Delta}(z,\hat{n})\rangle\nonumber \\&+ \langle \delta_{cv}(z^\prime) \tilde{\Delta} (z,\hat{n})\rangle \langle \delta_{cv}(z) \tilde{\Delta}(z^\prime,\hat{m})\rangle.\label{eq:expr2}
\end{align}
By integrating this equation over the Legendre polynomials we obtain $C_\ell$ on the left hand side. This gives us the constraint equation:
\begin{multline}\label{eq:crossterms}
    C_\ell(z,z^\prime) = \tilde{C}_\ell(z,z^\prime)\left(1+\frac{C_0(z,z^\prime)}{4\pi}\right) + C_0(z,z^\prime)\delta_{\ell,0} + \\  4\pi \left[\langle \tilde{\Delta}(z,\hat{n}) \delta_{cv}(z^\prime) \rangle + \langle \tilde{\Delta}(z^\prime,\hat{m}) \delta_{cv}(z) \rangle + \langle \delta_{cv}(z^\prime) \tilde{\Delta}(z^\prime ,\hat{m})\rangle \langle \delta_{cv}(z) \tilde{\Delta}(z,\hat{n})\rangle\right]\delta_{\ell,0}.
\end{multline}
The Kronecker delta arises when the Legendre polynomials are integrated over a constant. The $\langle \tilde{\Delta}\delta_{cv}\rangle$ correlators cannot depend on angle as $ \tilde{\Delta}(z,\hat{n})\delta_{cv}(z^\prime)$ only depends on one angle, which disappears due to the ensemble averaging.

Per definition $\tilde{C}_0=0$. Then, from considering $\ell=0$ we can deduce that the cross correlation terms have to be zero.  Without these terms we can rewrite (\ref{eq:expr2}) in terms of power spectra as:
\begin{equation}
    C_\ell(z,z^\prime) = \tilde{C}_\ell(z,z^\prime)\left(1+\frac{C_0(z,z^\prime)}{4\pi}\right)+C_0(z,z^\prime) \delta_{\ell,0}.
\end{equation}
The complete expression for $\tilde{C}_\ell(z,z^\prime)$ in terms of our predicted power spectrum $C_\ell(z,z^\prime)$ is then:
\begin{equation}\label{eq:angcl}
    \tilde{C}_\ell(z,z^\prime) = \frac{C_\ell(z,z^\prime) -C_0(z,z^\prime)\delta_{\ell,0}}{1+\frac{C_0(z,z^\prime)}{4\pi}}.
\end{equation}
From this we learn two things: First, indeed for $\ell=0$ this gives zero by construction.  Second, we can account for the different normalisation in the power spectrum by dividing (\ref{eq:clwrongnorm}) by a factor $1+\frac{C_0(z,z^\prime)}{4\pi}$. 

As an independent approach we can show that the $\langle \tilde{\Delta}\delta_{cv}\rangle$ correlator is zero when expanding in the cosmic variance. 
\begin{align}
    \langle \tilde{\Delta}(z,\hat{n}) \delta_{cv}(z^\prime) \rangle&=\left \langle \left(\frac{\hat{d}_L(z,\hat{n})}{\bar{d}_L(z)}-1\right) \delta_{cv}(z^\prime)\right \rangle \\
    & = \left \langle \left(\frac{\hat{d}_L(z,\hat{n})}{\langle{\hat{d}_L}\rangle(z)(1+\delta_{cv}(z))}+1\right) \delta_{cv}(z^\prime)\right \rangle\\
    & \approx \left \langle  \left(\frac{\hat{d}_L(z,\hat{n})}{\langle{\hat{d}_L}\rangle(z)}(1-\delta_{cv}(z))-1\right) \delta_{cv}(z^\prime)\right\rangle  +\mathcal{O}(\delta_{cv}^3).\label{eq:expansiondl}
\end{align}
Here, we expanded the cosmic variance up to linear order, assuming it is small. We can now rewrite (\ref{eq:expansiondl}) in terms of the tracers again,
\begin{align}
    \langle \tilde{\Delta}(z,\hat{n}) \delta_{cv}(z^\prime) \rangle   \approx& \left \langle \left(\frac{\hat{d}_L(z,\hat{n})}{\langle{\hat{d}_L}\rangle(z)}(1-\delta_{cv}(z))-1\right) \delta_{cv}(z^\prime)\right \rangle + \mathcal{O}(\delta_{cv}^3) \\ 
     =& \langle \Delta(z,\hat{n}) \delta_{cv}(z^\prime) \rangle - \langle \delta_{cv}(z) \delta_{cv}(z^\prime)\rangle\nonumber \\& + \langle \Delta(z,\hat{n})\delta_{cv}(z) \delta_{cv}(z^\prime)\rangle + \mathcal{O}(\delta_{cv}^3) .\label{eq:firstorder}
\end{align}
Now we drop the three point function as a consequence of their assumed Gaussianity. We can also derive an expression for the $\langle \Delta \delta_{cv} \rangle$ correlator by using the same trick as in deriving (\ref{eq:cvmono}). We factor out the angular integral out of the ensemble brackets to obtain:
\begin{align}
    \langle \Delta(z,\hat{n}) \delta_{cv}(z^\prime) \rangle &= \int \frac{d^2 \hat{m}}{4\pi}\left[ \frac{\langle \hat{d}_L(z,\hat{n}) \hat{d}_L(z^\prime,\hat{m}) \rangle}{\langle \hat{d}_L \rangle (z) \langle \hat{d}_L \rangle (z^\prime)}-1\right]\\ &= \frac{1}{2}\int_{-1}^1 d\mu\left[ \frac{\langle {d}_L(z) {d}_L(z^\prime) \rangle(\mu)}{\langle \hat{d}_L \rangle (z) \langle \hat{d}_L \rangle (z^\prime)}-1\right]\\ &= \frac{C_0(z,z^\prime)}{4\pi}.\label{eq:crosscor0}
\end{align}
This is precisely the same as (\ref{eq:cvmono1}). Inserting both (\ref{eq:cvmono1}) and (\ref{eq:crosscor0}) in (\ref{eq:firstorder}) then gives us:
\begin{equation}
    \langle \tilde{\Delta}(z,\hat{n}) \delta_{cv}(z^\prime) \rangle=0.
\end{equation}
Using this to simplify (\ref{eq:expr2}) and then rewriting it as power spectra then also results in (\ref{eq:angcl}). We note that we have now obtained that the angular averaged monopole is zero without demanding this. 
\section{Power spectra for Doppler and convergence effects}\label{app:dopplerlensing}
In this Appendix we derive the power spectra for the convergence and Doppler effects, discussed in Section \ref{sec:contaminants}. We derive these in the context of perturbed $\Lambda$CDM, where these fluctuations are well studied \cite{sasaki1987,Bonvin_2006,Biern_2017,Garoffolo2021}. We use these spectra as contaminant effects, which to be measurable our dark energy signal should rise above.
\subsection{Doppler}
For the Doppler effect we have (\ref{eq:dopp}): 
\begin{equation}
    \kappa_v(z,\hat{n})=\left(\frac{1}{\chi(z)\mathcal{H}(z)}-1\right)\vec{v}\cdot\hat{n}.
\end{equation}
Here, $\mathcal{H}=aH$ is the conformal Hubble parameter. We can now insert the Fourier transform of $\vec{v}(\hat{n},z)$ and use the continuity equation (\ref{eq:cont}) to obtain:
\begin{equation}
    \kappa_v(z,\hat{n}) = i \int\frac{d^3k}{(2\pi)^3}\left(\frac{1}{\chi(z)}-\mathcal{H}(z)\right)\frac{\vec{k}\cdot\hat{n}}{k^2}f(z)\delta(\vec{k},z)e^{i \vec{k}\cdot\hat{n}\chi(z)}.
\end{equation}
The factor $\vec{k}\cdot\hat{n}e^{i \vec{k}\cdot\hat{n}\chi(z)}$ can now be written using spherical harmonics to give us:
\begin{equation}
    \kappa_v(z,\hat{n}) = 4 \pi \sum_{\ell m} \int\frac{d^3k}{(2\pi)^3}\left(\frac{1}{\chi(z)}-\mathcal{H}(z)\right)f(z)\frac{\delta(\vec{k},z)}{k} i^{\ell+1}  j_{\ell}^{\prime}(k \chi(z)) Y_{\ell m}(\hat{k}) Y_{\ell m}(\hat{n}).
\end{equation}
Now, we can recognise this as an expansion in spherical harmonics, the expansion coefficients being given by
\begin{equation}\label{eq:kappav}
    \kappa_{v,\ell m} =  4 \pi i^{\ell+1}\left(\frac{1}{\chi(z)}-\mathcal{H}(z)\right)f(z) \int\frac{d^3k}{(2\pi)^3}\frac{\delta(\vec{k},z)}{k}  j_{\ell}^{\prime}(k \chi(z)) Y_{\ell m}(\hat{k}).
\end{equation}
The quantity of interest is the angular power spectrum $\langle \kappa_{v,\ell m}(z)\kappa^*_{v,\ell m}(z^\prime)\rangle = C^v_\ell(z,z^\prime)$. We relate this to the matter power spectrum, defined as:
\begin{equation}\label{eq:ps}
    \langle\delta(\vec{k},z)\delta^*(\vec{k}^\prime,z^\prime)\rangle = (2\pi)^3 P_m(k,z,z^\prime)\delta^{3}(\vec{k}-\vec{k^\prime}),
\end{equation}
where $P_m(k)$ is the matter power spectrum and $\delta^{3}$ is the three dimensional Dirac-delta function. Combining  (\ref{eq:kappav}) and (\ref{eq:ps}) we obtain
\begin{multline}
    C_\ell(z,z^\prime) =\frac{2}{\pi} \left(\frac{1}{\chi(z)}-\mathcal{H}(z)\right)f(z)\left(\frac{1}{\chi(z^\prime)}-\mathcal{H}(z)\right)f(z^\prime)\\ \times\int_0^\infty dk P_m(k) j^\prime_\ell(k\chi(z)) j^\prime_\ell(k\chi(z^\prime)).
\end{multline}
Integrating this over a redshift distribution $p(z)$ then gives us the following angular power spectrum:
\begin{multline}
    C^{v,i,j}_\ell=  \frac{2}{\pi}\int_0^\infty dz_1 p^i (z_1) \int_0^\infty dz_2 p^{\ j} (z_2)\left(\frac{1}{\chi(z_1)}-\mathcal{H}(z_1)\right)f(z_1)\left(\frac{1}{\chi(z_2)}-\mathcal{H}(z_2)\right)f(z_2) \\ \times\int_0^\infty dk P_m(k,z_1,z_2) j_{\ell}^{\prime}(k \chi(z_1))  j_{\ell}^{\prime}(k \chi(z_2)).
\end{multline}
\subsection{Convergence}
We can now go through the same treatment for the convergence term:
\begin{equation}
    \kappa_c(z,\hat{n})=\int_0^z \frac{dz^\prime}{H(z^\prime)}\frac{\chi(z)-\chi(z^\prime)}{\chi(z)\chi(z^\prime)}\Delta_\perp \left(\Phi(z,\vec{k})+\Psi(z,\vec{k})\right).
\end{equation}
$\Delta_\perp$ is the Laplacian transverse to  the line of sight and $\Phi$ and $\Psi$ the Bardeen potentials. Again, we start with going to Fourier space and writing the exponent in spherical harmonics. We can use that the spherical harmonics are an eigenbasis of $\Delta_\perp$ with eigenvalues $\ell(\ell+1)$. We then obtain:
\begin{multline}
    \kappa_c(z,\hat{n}) = 4 \pi \sum_{\ell m} \ell(\ell+1) \int\frac{d^3k}{(2\pi)^3}\int_0^z \frac{dz^\prime}{H(z^\prime)}\left\{\frac{\chi(z)-\chi(z^\prime)}{\chi(z)\chi(z^\prime)}\right.\\\left. \times\left(\Phi(z,\vec{k})+\Psi(z,\vec{k})\right) i^{\ell}  j_{\ell}(k \chi) Y_{\ell m}(\hat{k}) Y_{\ell m}(\hat{n})\right\}.
\end{multline}
The coefficients of the spherical harmonics expansion can now seen to be:
\begin{multline}\label{eq:trac}
    \kappa_{c,\ell m}(z) =4\pi\ell(\ell+1) \int\frac{d^3k}{(2\pi)^3}\int_0^z \frac{dz^\prime}{H(z^\prime)}\left\{\frac{\chi(z)-\chi(z^\prime)}{\chi(z)\chi(z^\prime)}\right. \\ \left. \times\left(\Phi(z,\vec{k})+\Psi(z,\vec{k})\right) i^{\ell}  j_{\ell}(k \chi(z)) Y_{\ell m}(\hat{k})\right\}.
\end{multline}
We want to make the connection to the matter power spectrum $P_m(k)$, therefore we need to connect the Bardeen potentials to the overdensity $\delta(k,z)$. This is done with the following transfer function, which can be derived from the Poisson equation \cite{dodelson2020modern},
\begin{equation}\label{eq:poisson}
T_{\Phi+\Psi}(k,z) =  -\frac{3 H_0^2 \Omega_{M}}{k^2 a(z)}.
\end{equation}
 We now have all the ingredients to calculate the desired angular power spectrum. Combining  (\ref{eq:ps}), (\ref{eq:trac}) and (\ref{eq:poisson}) we obtain:
\begin{multline}
    C_\ell(z,z^\prime) = \frac{2}{\pi}(\ell(\ell+1))^2\int_0^z \frac{dz_1}{H(z_1)}\frac{\chi(z)-\chi(z_1)}{\chi(z)\chi(z_1)}\int_0^{z^\prime} \frac{dz_2}{H(z_2)}\frac{\chi(z^\prime)-\chi(z_2)}{\chi(z^\prime)\chi(z_2)}\\ \times \int_0^\infty dk k^{2} T_{\Phi+\Psi}(k,z_1) T_{\Phi+\Psi}(k,z_2) P_m(k,z_1,z_2) j_\ell(k\chi(z_1)) j_\ell(k\chi(z_2)). 
\end{multline}
This can now be integrated over some redshift distribution. Then, the order of integration can be changed in the following way: $\int_0^\infty dz \int_0^z dz^\prime \rightarrow \int_0^\infty dz^\prime \int_{z^\prime}^\infty dz $. This then gives the total power spectrum:
\begin{multline}
    C^{c,i,j}_\ell = \frac{2}{\pi}(\ell(\ell+1))^2 \int_0^\infty  \frac{dz_1}{H(z_1)}W^{\ i}_L(z_1)\int_0^{\infty} \frac{dz_2}{H(z_2)}W^{\ j}_L(z_2)\\ \times \int_0^\infty dk k^{2} T_{\Phi+\Psi}(k,z_1) T_{\Phi+\Psi}(k,z_2) P_m(k,z_1,z_2) j_\ell(k\chi(z_1)) j_\ell(k\chi(z_2)). 
\end{multline}
With $W^i_L(z_i)$ being the lensing kernel defined as:
\begin{equation}
    W^i_L(z_i) = \int_{z_i}^\infty dz p^i (z) \frac{\chi(z)-\chi(z_i)}{\chi(z)\chi(z_i)}.
\end{equation}




\bibliographystyle{unsrt}
\bibliography{biblio}
\end{document}